\PassOptionsToPackage{svgnames}{xcolor}
\documentclass[twocolumn]{aastex63}
\usepackage{fixfoot}
\usepackage[clockwise, figuresright]{rotating}
\usepackage{amsmath}
\usepackage{gensymb}
\usepackage{color,soul}
\usepackage{multirow}
\usepackage{IEEEtrantools}
\usepackage{bm}
\usepackage{float}
\usepackage{textcomp}
\usepackage{booktabs}
\usepackage{longtable}
\usepackage{csquotes}
\usepackage{morefloats}
\usepackage{scalerel}
\usepackage{appendix}
\published{2020 November 06 in the Astrophysical Journal}

\shorttitle{The  M--$\rm n_{sph}$ and M--$\rm R_{e,sph}$ relations}
\shortauthors{Sahu, Graham, and Davis}

\begin{document}
\title{Defining the (Black Hole)-Spheroid Connection with the Discovery of Morphology-Dependent Substructure in the $M_{\rm BH}$--$\rm n_{sph}$ and $M_{\rm BH}$--$\rm R_{e, sph}$ Diagrams: New Tests for Advanced Theories and Realistic Simulations}

\correspondingauthor{Nandini Sahu}
\email{nsahu@swin.edu.au}

\author[0000-0003-0234-6585]{Nandini Sahu}
\affil{OzGrav-Swinburne, Centre for Astrophysics and Supercomputing, Swinburne University of Technology, Hawthorn, VIC 3122, Australia}
\affil{Centre for Astrophysics and Supercomputing, Swinburne University of Technology, Hawthorn, VIC 3122, Australia}

\author[0000-0002-6496-9414]{Alister W. Graham}
\affil{Centre for Astrophysics and Supercomputing, Swinburne University of Technology, Hawthorn, VIC 3122, Australia}

\author[0000-0002-4306-5950]{Benjamin L. Davis}
\affil{Centre for Astrophysics and Supercomputing, Swinburne University of Technology, Hawthorn, VIC 3122, Australia}

\nocollaboration{3}


\begin{abstract}
For 123 local galaxies with directly-measured black hole masses ($M_{\rm BH}$), we provide the host spheroid's S\'ersic index ($\rm n_{sph}$), effective half-light radius ($\rm R_{e,sph}$), and effective surface brightness ($\mu_e$),  obtained from careful multi-component decompositions, and we use these to derive the morphology-dependent $M_{\rm BH}$--$\rm n_{sph}$ and $M_{\rm BH}$--$\rm R_{e,sph}$ relations. We additionally present the morphology-dependent $M_{\rm *,sph}$--$\rm n_{sph}$ and $M_{\rm *,sph}$--$\rm R_{e,sph}$ relations. We explored differences due to: early-type galaxies (ETGs) versus late-type galaxies (LTGs); S\'ersic versus core-S\'ersic galaxies; barred versus non-barred galaxies; and galaxies with and without a stellar disk. We detect two different $M_{\rm BH}$--$\rm n_{sph}$ relations due to ETGs and LTGs with power-law slopes $3.95\pm0.34$ and $2.85\pm 0.31$. We additionally quantified the correlation between $M_{\rm BH}$ and the spheroid's central concentration index, which varies monotonically with the S\'ersic index. Furthermore, we observe a single, near-linear $M_{\rm *,sph}$--$\rm R_{e,sph}^{1.08\pm 0.04}$ relation for ETGs and LTGs, which encompasses both classical and alleged pseudobulges. In contrast, ETGs and LTGs define two distinct $M_{\rm BH}$--$\rm R_{e,sph}$ relations with $\Delta_{\rm rms|BH}\sim\rm 0.60~dex$ (cf.\ $\sim$0.51~dex for the $M_{\rm BH}$--$\sigma$ relation and $\sim$0.58~dex for the $M_{\rm BH}$--$M_{\rm *,sph}$ relation), and the ETGs alone define two steeper $M_{\rm BH}$--$\rm R_{e,sph}$ relations, offset by $\sim$1~dex in the $\log M_{\rm BH}$-direction, depending on whether they have a disk or not and explaining their  similar offset in the $M_{\rm BH}$--$M_{\rm *,sph}$ diagram. This trend holds using $10 \%$, $50 \%$, or $90 \%$ radii.  These relations offer pivotal checks for simulations trying to reproduce realistic galaxies, and for theoretical studies investigating the dependency of black hole mass on basic spheroid properties.
\end{abstract}

\section{\textbf{Introduction}}

It is widely known that the mass of the black hole (BH) residing at the centre of most galaxies is correlated with both the host spheroid's stellar mass ($M_{\rm *,sph}$) and its central stellar velocity dispersion ($\sigma$). At the same time, bulgeless galaxies, for example, NGC~2478, NGC~4395, and NGC~6926,  have also been observed to house massive BHs \citep[e.g.][and references therein]{Secrest:Satyapal:2013, Simmons2013,denBrok:2015,Davis:2018:a}, and one of the tightest scaling relations is between black hole mass ($M_{\rm BH}$) and the winding/pitch angle of the spiral arms in spiral galaxies \citep{SeigarKennefick2008, Berrier:2013,Davis:Graham:2017}.  Additional correlations exist between $M_{\rm BH}$ and disk stellar mass \citep{Davis:2018:b}, disk rotation, and dark matter halo mass \citep{Ferrarese2002, Baes:Buyle:2003, Volonteri:Vrot:2011, Davis:Graham:Combes:2019}. 
Collectively, this goes beyond the notion of a single primary (causal) relation for all galaxies plus secondary (indirect/consequential) relations, and reveals a greater level of complexity. 
Indeed, the markedly different $M_{\rm BH}$--$M_{\rm *,gal}$ and $M_{\rm BH}$--$M_{\rm *,sph}$ relations for early-type galaxies (ETGs, comprised of E-, ES\footnote{ES-type represents ellicular galaxies which have an intermediate-scale stellar disk confined to within their spheroid \citep{Liller:1966, Graham:Grid:2019}.}-, and S0-types) and late-type galaxies (LTGs), i.e.\ spiral (Sp) galaxies \citep{Davis:2018:b, Davis:2018:a,Sahu:2019:I},   undoubtedly reflects the different physical processes, gas supply history, net angular momentum, involved in building these systems.  

The review of the BH scaling relations by \citet{Graham:2016:Review} highlighted the need to achieve internal consistency among the various scaling relations, in particular between the $M_{\rm BH}$--$\sigma$, $M_{\rm BH}$--$M_{\rm *,sph}$, and $\sigma$--$M_{\rm *,sph}$ relations. This followed \citet{Graham:2012} who reported on a near-linear and super-quadratic $M_{\rm BH}$--$M_{\rm *,sph}$ relation, respectively, for spheroids with a S\'ersic or core-S\'ersic\footnote{Core-S\'ersic galaxies have a deficit of light at their centre; hence,  their central (bulge) light profile is described using a shallow power-law followed by a \citet{Sersic:1963} function beyond the core \citep{Graham:2003:CS}. This population was first discussed by \citet{King:Minkowski:1966, King:Minkowski:1972}.}  light profile (see also \citet{Graham:Scott:2013} and \citet{ Scott:Graham:2013}).
\citet{Savorgnan:2016:Slopes} subsequently discovered an improved division due to ETGs  and LTGs (none of LTGs have core-S\'ersic bulge profiles) in the $M_{\rm BH}$--$M_{\rm *,sph}$ diagram, and in the $M_{\rm BH}$--($\rm L_{gal}$, galaxy luminosity) diagram. This was also later  reported by \citet{van_den_Bosch:2016}.  \citet{Savorgnan:2016:Slopes} coined the notion of a red and blue sequence when two tracks, due to ETGs and LTGs, are evident in a BH mass scaling diagram \citep[see also][]{Terrazas:Bell:2016, Dullo:Bouquin:2020}.
\citet{Sahu:2019:I} additionally found that the $M_{\rm BH}$--$M_{\rm *,sph}$ relation for ETGs with a disk (ES and S0) and  ETGs without a disk (E-type) is roughly quadratic, while the two relations are offset by more than an order of magnitude in the $M_{\rm BH}$-direction. This has since been found in a recent simulation by \citet{Marshall2020}. Clearly, it is not simply the amount of stellar mass that matters, but also how it was accumulated and is now distributed.  In this vein, we explore the relationship that the BH mass has with the size and shape (centrally concentrated or diffused) of the surrounding bulge/spheroid--- terms that we use interchangeably---and as a function of the morphology of the host galaxy. 

The above mentioned developments  represent a key advance in our understanding of the coevolution of galaxies and black holes.  It built upon works such as \citet{Wandel:1999}, \citet{Salucci:2000}, \citet{Laor:2001}, and \citet{Graham:2012}  and voided the notion \citep{Dressler:1989,Kormendy:Richstone:1995,Magorrian:1998} that the black hole mass simply co-evolved linearly with the spheroid mass.  The recognition of a more nuanced situation is perhaps not surprising given the variety of accretion/merger histories, and resulting structures among galaxies. For example, core-S\'ersic galaxies, thought to be built from the dry merger of galaxies with pre-existing black holes \citep{Begelman1984}, appear to follow a steeper relation in the $M_{\rm BH}$--$\sigma$ diagram \citep{Sahu:2019:II}, see also \citet[][their Figure 3a]{Terrazas:Bell:2016} and \citet[][their Figure 5]{Bogdan:2018}. 

Based on the low intrinsic scatter about the $M_{\rm BH}$--$\sigma$ relation \citep{Ferrarese:Merritt:2000,Gebhardt:2000}, some studies have concluded that it is the most fundamental relation between black hole mass and galaxy \citep[e.g.][]{van_den_Bosch:2016, Nicola:2019}. However,  it may be a premature conclusion without considering the correlations between BH mass and other basic galaxy properties, or allowing for the morphology-dependence and thus (formation physics)-dependence of galaxies. Moreover, it overlooks that the $M_{BH}$--pitch angle ($\phi$) relation has the least total scatter at 0.43~dex \citep{Davis:Graham:2017} compared to 0.51~dex in the latest $M_{\rm BH}$--$\sigma$ diagram \citep{Sahu:2019:II}.

Establishing if a, and which, single relation is the most fundamental, i.e., the primary relation, and how it depends upon morphology is important for understanding the co-evolution of galaxies and BHs. The secondary scaling relations --- not to be confused with the morphology dependent substructure which reveals an additional parameter/factor modulating the co-joined growth of galaxies and BHs\footnote{We could re-frame these results by constructing a simplified `fundamental plane', i.e.\ a 3-parameter equation involving $M_{\rm BH}$, $\sigma$ (or $M_*$) and morphological type (even if just a binary parameter). This would effectively unite the morphology-dependent $M_{\rm BH}$--$\sigma$ ($M_{\rm BH}$--$M_{*}$)  relations and reduce the scatter about the two-parameter relations which ignore the morphological type. We will pursue this in future work.}  --- are, however, also important. They can still be used, for example, to predict BH masses or to check on the accuracy of computer simulations e.g., CLUES \citep{Yepes:CLUES:2009}, Magneticum \citep{Dolag:Magneticum:2015}, Bolshoi \citep{Klypin:Bolshoi:2011}, EAGLE \citep{SchayeEagle2015}, Illustris \citep{Vogelsberger:Illustris:2014}, IllustrisTNG \citep{Pillepich:IllustrisTNG:2018}, FIRE \citep{HopkinsFIRE2018}, and SIMBA \citep{Dave:Simba:2019}, which are trying to produce realistic galaxies\footnote{Simulations lacking primary information about the spheroid can still be tested against the non-linear, morphology-dependent, $M_{\rm BH}$--$M_{\rm *, gal}$  relations \citep{Davis:2018:b, Sahu:2019:I}.}.  These empirical relations help to decipher the physics behind the effect of the central supermassive black hole on the host spheroid/galaxy properties and vice versa. How such
black hole feedback drives galaxy evolution is the challenge yet to be fully answered \citep{Choi:Somerville:2018, Ruszkowski2019, TerrazasIllustrisTNG2020, MartinNavarro2020}.

Here, we will expand upon the previous efforts in establishing the $M_{\rm BH}$--$\rm n_{sph}$ relation \citep[e.g.,][]{GrahamErwin2003, Graham:Driver:2007, Vika:Driver:2012, Beifiori:Courteau:2012, Savorgnan2013,  Savorgnan:n:2016}, the $M_{\rm *, sph}$--$\rm n_{sph}$ relation  \citep[e.g.,][]{Andredakis1995, Jerjen2000, Graham:Guzman:2003, FerrareseCote2006, Savorgnan:n:2016}, the $M_{\rm *, sph}$--$\rm R_{e,sph}$ relation \citep[e.g.,][]{Sersic:1968, Graham:Worley:2008, Lange2015}, and  the $M_{\rm BH}$--$\rm R_{e,sph}$ relation \citep[e.g.,][]{Nicola:2019} using our extensive sample of 83 ETGs and 40 LTGs with careful (individual, not automated)  multi-component decompositions. Importantly, we explore potential substructures due to galaxy sub-morphologies, i.e., S\'ersic versus core-S\'ersic galaxies, barred versus non-barred galaxies, galaxies with a stellar disk versus galaxies without a stellar disk, and ETGs versus LTGs.  We also investigate the relation between $M_{\rm BH}$ and the central concentration index \citep{GrahamErwinCaon2001}, which is known to vary  monotonically with the S\'ersic index \citep{Trujillo:Graham:Caon:2001, GrahamTrujilloCaon2001}. 
 
As with the $M_{\rm BH}$--$M_{\rm *,sph}$ relation, the $M_{\rm BH}$--$\rm n_{sph}$ and  $M_{\rm BH}$--$\rm R_{e,sph}$ relations can be applied to large surveys of galaxies \citep[e.g.,][]{Casura:GAMA:2019} --- if their bulge S\'ersic parameters are reliable  ---  to estimate their black hole masses and further construct the black hole mass function (BHMF).  The BHMF holds interesting information for cosmologists, e.g., to estimate the mass density of the Universe contained in BHs \citep[e.g.][]{Fukugita:Peebles:2004, GrahamDriver2007BHMD}, to map the growth of BHs and constrain theoretical models of BH evolution  \citep[e.g.][]{Kelly2012}.  In addition, the latest BHMF, which takes part in calculating the black hole merger rate \citep{Chen2019,Volonteri:Pfister:2020}, will help improve the prediction for the amplitude and time until detection of the long-wavelength (micro to nano Hertz)  gravitation wave background --- as generated from merging supermassive black holes --- using pulsar timing arrays \citep{Siemens2013, Shannon:2015, Sesana2016} and using the upcoming Laser Interferometer Space Antenna \citep[LISA,][]{Danzmann:2017, Baker:2019}. 

Section \ref{Data} details the galaxy sample and parameters which we used for our investigation, and the regression routines applied to obtain the correlations. Various correlations we observed, including their dependencies on galaxy morphology, are described in the subsections of Section \ref{results}. In sub-section \ref{3.1}, we present the scaling relations observed between the spheroid stellar mass and spheroid S\'ersic index. Sub-section \ref{3.2} presents the expected correlation between black hole mass and the bulge S\'ersic index by combing the correlation observed between spheroid stellar mass and spheroid S\'ersic index with our latest correlation between black hole mass and spheroid stellar mass. It then  presents the observed correlations between black hole mass and the bulge S\'ersic index based on our data-set. We also show the relationship between the S\'ersic index and the central light concentration, and we present the correlation observed between the black hole mass and the central concentration index. In sub-section \ref{3.3}, we present the correlations observed between the spheroid stellar mass and the effective spheroid half-light radius. Here, we also explore the correlations of the spheroid stellar mass with the spheroid radii containing $\rm 10 \%$ and $\rm 90 \%$ of the light of the spheroid. Sub-section \ref{3.4} provides the expected correlation that the spheroid half-light radius might have with the black hole mass, before presenting the observed correlations between the black hole mass and the spheroid effective half-light radius, along with the correlations between the black hole mass and the spheroid radii containing $\rm 10 \%$ and $\rm 90 \%$ of spheroid's light. These subsections additionally provide a discussion and some explanation for the correlations that we find. Finally, Section \ref{summary} presents a summary of our main results.

\section{\textbf{Data}}
\label{Data}
The \citet{Sersic:1963, Sersic:1968} function is nowadays used to describe the light profiles of elliptical galaxies (E) and, when present, the spheroidal component of galaxies with a disk (ES/S0/Sp). A review of the S\'ersic function, and its many associated expressions, can be found in \citet{Graham:Driver:2005}. 
Briefly, the intensity of a S\'ersic light profile can be  expressed as a function of the projected galactic radius (R), such that
 \begin{IEEEeqnarray}{rCl}
\label{Sersic}
\rm I(R)= I_e \exp{\left[-b_n \left \{ \left(\frac{R}{R_e} \right)^{1/n}-1 \right \}\right]},
\end{IEEEeqnarray}
 where $\rm I_e$, $\rm R_e$, and $\rm n$ are profile parameters. The term $\rm I_e$ is the effective intensity at the effective radius $\rm R_{e}$, which bounds 50\% of the total light in the associated 2D image. \citet{Graham:Re:2019} provides a detailed review of this popular radius and addresses the misconceptions about its physical significance. The surface brightness at this effective radius ($\rm \mu_e$) is related to $\rm I_e$  through $\rm \mu_e  \equiv  -2.5\log(I_e)$.

The S\'ersic index n (also known as the shape parameter), describes the curvature of the light profile, such that a S\'ersic light profile with a higher S\'ersic index is steeper at the centre and has a shallower distribution at larger radius, whereas, a profile with a smaller S\'ersic index  is shallower at the centre followed by a steeper drop at outer radii \citep[see Figure 2 in][]{Graham:Re:2019}. Thus, the S\'ersic index traces the central concentration of the light within the spheroid \citep[][their Figure 2]{Trujillo:Graham:Caon:2001, GrahamTrujilloCaon2001}; and also the inner gradient of the gravitational potential\footnote{This holds when dark matter is negligible, and there is no significant stellar mass-to-light ratio gradient.}  \citep[][their Figures 2 and 3]{Terzic:Graham:2005}. 
The value of the term $\rm b_n$ in Equation \ref{Sersic} depends on n, and is obtained by solving $\Gamma (\rm 2n) = 2 \gamma (\rm 2n, b_n) $, where $\Gamma$ denotes the gamma function and $\gamma$ is the incomplete gamma function.  It can also be approximated by $\rm b_n \approx 1.9992\, n - 0.3271 $ for $0.5 < n < 10$ \citep{Capaccioli:1989}.  

In this work, we have used a sample of 123 galaxies with directly-measured black hole masses, for whom the S\'ersic model parameters ($\rm n, R_e, and \, \mu_e$) describing their spheroid's surface brightness distribution were obtained by a careful multi-component decomposition of the galaxy's light. 
 These parameters are collectively taken from \citet{Savorgnan:Graham:2016:I}, \citet{Davis:2018:a}, and \citet{Sahu:2019:I}. These studies performed a 2-dimensional (2D) isophotal analysis, first extracting a 2D luminosity model using \textsc{Isofit} and \textsc{Cmodel} \citep{Ciambur:2015:Ellipse} to capture the radial gradients in the ellipticity, position angle, and Fourier harmonic coefficients describing the isophote's deviations from a pure ellipse, and then performing a  multi-component decomposition using the isophotal-averaged 1D surface-brightness profile along the major and geometric-mean\footnote{The geometric-mean axis, which is also known as the \enquote{equivalent axis}, is the radius of the circularized form of the elliptical isophote with major axis radius  $\rm R_{ maj}$ and minor axis radius $\rm R_{ min}$, which conserves the same amount of flux. This results in the equivalent axis radius ($R_{\rm eq}$) being the geometric mean of $\rm R_{ maj}$ and $\rm R_{ min}$ ($\rm R_{eq}=\sqrt{\rm R_{ maj}*R_{min}}$), which is also represented as $\rm R_{geom}$ \citep[for more details see the appendix section in][]{Ciambur:2015:Ellipse}.} axis of the galaxies. For this purpose, they used the software \textsc{Profiler} \citep{Ciambur:2016:Profiler}, which is inbuilt with many functions for specific galaxy components, including the S\'ersic function for galactic spheroids. The major and geometric-mean axis were modelled independently \citep[see Section 3 in][for more details]{Sahu:2019:I}. 

\begin{figure}
\begin{center}
\includegraphics[clip=true,trim= 12mm 08mm 21mm 21mm,width=   1.0\columnwidth]{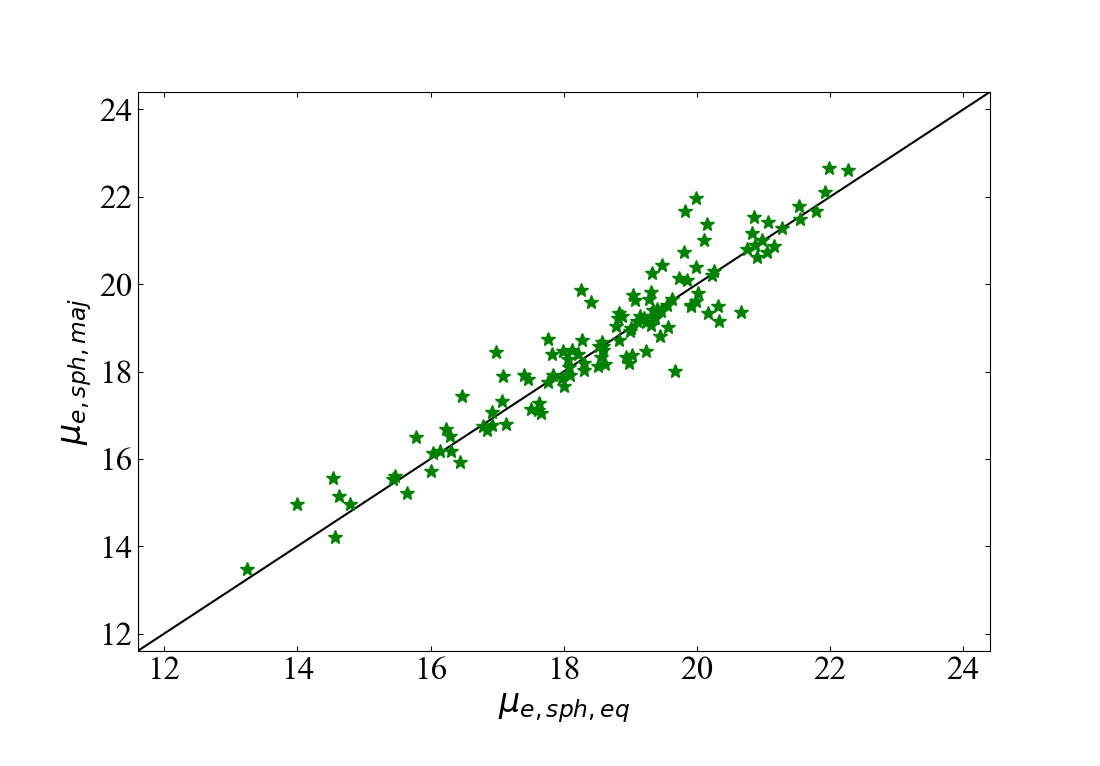}
\caption{The spheroid surface brightness at the effective half-light radius from a S\'ersic fit to the  major-axis light profile ($\mu_{\rm e,sph,maj}$) plotted against the spheroid surface brightness at the effective half-light radius from a S\'ersic fit to the  the geometric mean-axis  light profile ($\mu_{\rm e,sph,eq}$). This tight distribution of data-points over the one-to-one line demonstrates the consistency between the two independent decompositions.}
\label{mu_mu}
\end{center}
\end{figure}

Table \ref{Total Sample} in our appendix lists both the major-axis bulge parameters ($\rm n_{maj}, \, R_{e, maj}, \, \mu_{e, maj}$), and the equivalent-axis bulge parameters ($\rm n_{eq}, \, R_{e, eq}, \, \mu_{e, eq}$), plus the morphologies, and the bulge masses ($M_{*,\rm sph}$) taken from \citet{Savorgnan:Graham:2016:I}, \citet{Davis:2018:a}, and  \citet{Sahu:2019:I},  along with the distances and the directly-measured black hole masses of the galaxies. To show the consistency between the structural decomposition of the major- and equivalent-axis surface brightness profiles, we have plotted $\rm \mu_{e,sph, maj}$ versus $\rm \mu_{e, sph,eq}$  in Figure \ref{mu_mu}.  The 1$\sigma$ scatter in this diagram is $\rm 0.58 \, mag \, arcsec^{-2}$ which corresponds to a 1$\sigma$ scatter in $\rm R_{e}$ of $\approx$30\% given that the S\'ersic model's surface brightness profile has slopes of $\sim$1.8 to $\sim$2.1 (for $\rm n=1$ to 10) at $\rm R=R_{\rm e}$, where $\rm {\rm d} \mu (R)/{\rm dR}|_{R_{e}} = 2.5b_n/(\ln(10)\, n \, R_e) \approx [2.17-0.36/n] /Re$.
Table \ref{Total Sample} also provides the radial concentration index (C: see Section \ref{results}) and the physical (arcsec to kpc) size scale of the galaxies\footnote{The physical scale is calculated using the python version of Edward (Ned) L. Wright's cosmological calculator \citep{Wright2006}, written by James Schombert, assuming the cosmological parameters $\rm H_0=67.4 \, (km \, s^{-1})/Mpc$, $\Omega_m$=0.315, and $\Omega_v$=0.685 \citep{Planck:Collaboration:2018}.}. The morphologies of these galaxies are based on the  multi-component decompositions found in \citet{Savorgnan:Graham:2016:I}, \citet{Davis:2018:a}, and \citet{Sahu:2019:I}. 

The black hole masses used here have been obtained from various sources in the literature. Their original sources are listed in \citet{Savorgnan:2016:Slopes} and \citet{Sahu:2019:I} for the ETGs, and in  \citet{Davis:2018:a} for the LTGs. These black hole masses have been directly-measured using either the stellar dynamical modelling, gas dynamical modelling, megamaser kinematics, proper motions (Sgr $A^*$), or the latest direct imaging method (M87*). As the distances to the galaxies have been revised over time, the BH masses have also been updated to keep pace with this, and thereby provide a consistent analysis with the arcsecond-to-kpc and apparent-to-absolute magnitude conversions.

Our total sample is comprised of 123 galaxies, of which 83 are ETGs, and 40 are LTGs. We have used the  Bivariate Correlated Errors and Intrinsic Scatter (\textsc{BCES}) regression \citep{Akritas:Bershady:1996} to obtain the symmetric (bisector) best-fit lines for all our correlations. The \textsc{BCES}\footnote{We used the Python module from \citep{Nemmen:2012}, which is available at \url{https://github.com/rsnemmen/BCES}} regression considers the measurement errors in both  variables and allows for intrinsic scatter in the data. It is a modified form of the ordinary least square (OLS) regression. It calculates the  \textsc{OLS($Y|X$)} line by minimizing the scatter in the Y-direction,  and the \textsc{OLS($X|Y$)} line by minimizing the scatter in the X-direction. The  \textsc{BCES(Bisector)} line  symmetrically bisects the \textsc{OLS($Y|X$)} and \textsc{OLS($X|Y$)} lines. We prefer to use the bisector line as it offers equal treatment to the quantities plotted on the X-and Y-axes. Additionally, we also checked the consistency of our correlations by employing the modified-\textsc{FITEXY}  (\textsc{MPFITEXY}) regression \citep{Press:1992, Tremaine:ngc4742:2002, Williams:2010, Markwardt:2012}, where we had to bisect the best-fit lines obtained from the forward \textsc{MPFITEXY($Y|X$)} and inverse \textsc{MPFITEXY($X|Y$)}  regressions to obtain the symmetric fit to our data \citep[see][for more details about the  \textsc{MPFITEXY} regression]{Novak:2006}.

For our investigation, we adopt a $20\%$ uncertainty for the  S\'ersic bulge parameter n. Various factors which can contribute to the uncertainty in the measurement of the S\'ersic  bulge parameters include: inappropriate sky subtraction; incomplete masking; inaccurate point-spread function (PSF) for the telescope; uncertainties in the identification of components; especially the nuclear (bar/disk/ring/star cluster)  or faint components during the multi-component decomposition of the galaxy luminosity. Thus, it is challenging to quantify the uncertainty in the bulge parameters for every galaxy individually. 

In past studies, various measures have been taken to quantify realistic errors on  the bulge/galaxy S\'ersic index. For example, \citet{Caon1993} noted a typical error of $\sim 25 \%$ corresponding to a $25\%$ variation in the (observed - fitted) residual, while some studies  \citep[e.g.][]{ Graham:Driver:2007, Savorgnan2013} adopted a constant uncertainty of $\sim 20 \%$, and others employed Monte Carlo simulations (e.g. \citet{Beifiori:Courteau:2012} obtaining up to a $\sim 15\%$ error-bar). Others varied the sky subtraction by $\pm 1 \sigma$ to estimate error-bars \citep{Vika:Driver:2012}, some used mean/median errors based on a broader comparison with published parameters from other studies  \citep{Graham:Worley:2008, Laurikainen:Salo:2010} producing up to $\sim 30\%$ uncertainty, whereas \citet{Savorgnan:n:2016} used $20\%$, $42\%$, and $52\% $ uncertainties, respectively, for their grade 1, grade 2, and grade 3 galaxies following \citet[][their Section 4.2]{Savorgnan:Graham:2016:I}. As \citet{Savorgnan:Graham:2016:I} noted, their generous uncertainties arose when comparing published parameters based upon an array of differing decompositions for the same galaxy. For example, sometimes a single S\'ersic component had been fit while other times the image analysis additionally included, as separate components, a disk and sometimes also a bar.

Given that our sky-background intensities are measured carefully \citep[][see their Figure 1 and Section 2.2 ]{Sahu:2019:I}, and that our parameters are obtained from multi-component decompositions, we have ruled out our two major sources of systematic errors  (i.e.\ over/under-estimation of the sky and failing to account for a biasing component), and as such we adopt a 20\% uncertainty for n, and a 30\% uncertainty for $\rm R_e$ based on the $1 \sigma$ scatter in $\rm \mu_e$ for our galaxy sample as already described in this section. We do, however, test and confirm that our scaling relations are not significantly dependent upon this.  Our results are stable (no change in slope or intercept at the $1\sigma$ uncertainty level) upon using an uncertainty up to 30\% in $\rm n$ and 40\% in $\rm R_e$. Furthermore, we also performed all the correlations using the major subsample of our total sample for whom the spheroid parameters are derived using $\rm 3.6 \,\mu m$ images (see Table \ref{Total Sample}), and the correlations are found to be consistent with the correlations obtained using the total sample within the $\pm 1 \sigma$ uncertainty bounds of the slopes and intercepts.

During our linear regressions, we have excluded certain potentially biasing galaxies, which are either stripped galaxies (NGC~4342 and NGC~4486B), a single galaxy with $ M_{\rm BH} < 10^5 M_{\odot}$ (NGC~404), or more than $2\sigma$ outliers (NGC~1300, NGC~3377, NGC~3998, NGC~4945, NGC~5419) in any of the correlations presented here.  NGC~4342 and NGC~4486B are stripped of their stellar mass due to the gravitational pull of their massive companion galaxies NGC~4365 \citep{Blom:Forbes:2014} and NGC~4486 \citep{Batcheldor:2010:b}, respectively. Hence, NGC~4342 and NGC~4486B can bias the black hole scaling relations as they have smaller $\rm n$ or $\rm R_e$ than they would have had if they weren't stripped of their mass. NGC~404,  the only galaxy in our sample with a BH mass below $10^6 M_{\odot}$, can bias the best-fit lines due to its location at the end of the distribution and thus its elevated torque strength. The galaxies NGC~3377, NGC~3998, NGC~4945, and NGC~5419 in the $M_{\rm BH}$--$\rm n$ diagram, and NGC~1300 in the $M_{\rm BH}$--$\rm R_e$ diagram, are more than $\pm 2\sigma_{\rm rms}$ outliers from the corresponding best-fit lines and slightly alter their slopes\footnote{Including these galaxies in the regressions changes the slopes by $\sim1\sigma$ uncertainty level of current slopes.}. Hence, these galaxies are better excluded in all our regressions to obtain robust correlations. These eight excluded galaxies are indicated in all the plots. This exclusion leaves us with a reduced sample of 115 galaxies.
   
\section{\textbf{Scaling relations}}
\label{results}
The stellar masses of our galactic spheroids  ($M_{\rm *,sph}$) are derived from the luminosities measured using the S\'ersic model (for the bulge) fit to the equivalent- (or geometric-mean) axis light profile, parameterized by $\rm n_{sph,eq}$, $\rm R_{e,sph,eq}$, and $\rm I_{e,sph,eq}$. Therefore, it is expected to find some correlation between $M_{\rm *,sph}$ and the S\'ersic index, and also between $M_{\rm *,sph}$ and the effective half-light radius. The issue of parameter coupling potentially explaining the trends between the S\'ersic parameters and the  luminosity was explored and dismissed using model-independent measures of both luminosity and size \citep{Caon:Capaccioli:1993, Trujillo:Graham:Caon:2001}, implying the observed correlation between luminosity versus S\'ersic properties (n and $\rm R_e$)  are indeed real. Moreover, the  errors in n and $\rm R_e $ adopted here are not big enough for parameter coupling in the fitting process to explain the observed trends.

\subsection{\textbf{The} \boldmath{$M_{\rm *, sph}- \rm n_{sph}$} \textbf{diagram}}
\label{3.1}

\begin{figure*}
\begin{center}
\includegraphics[clip=true,trim= 07mm 03mm 12mm 12mm,width=   1.0\textwidth]{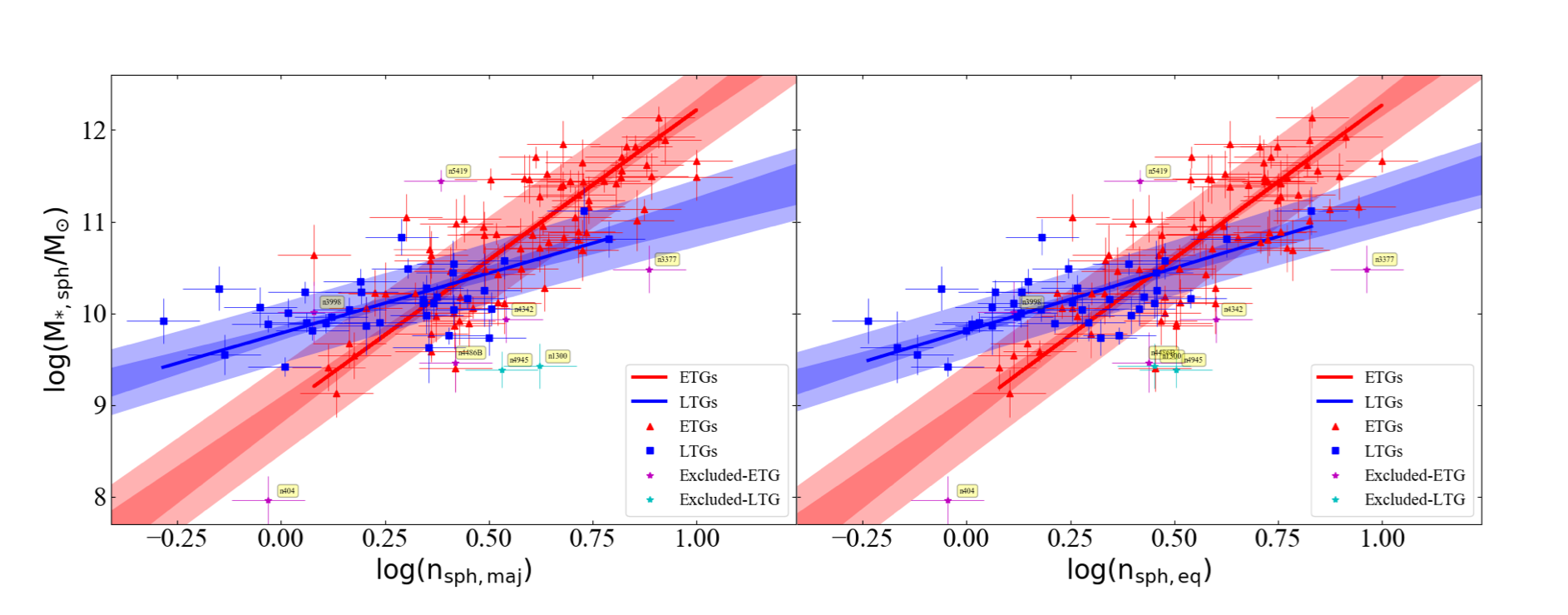}
\caption{Spheroid mass versus major-axis (left panel) and equivalent-axis (right panel) S\'ersic index describing the bulge/spheroidal component of the galaxies. In both panels, ETGs and LTGs are represented in red and blue, respectively. The bold red line for ETGs and blue line for LTGs represent the (symmetric) best-fit relations obtained using the \textsc{BCES(Bisector)} regression. The dark shaded region around these lines represents the $\pm 1\sigma$ uncertainty bound on the slopes and intercepts of these lines. The light-shaded region about these lines represent the $\pm 1 \sigma$ scatter bound of the corresponding dataset.
Both panels display the different $M_{\rm *, sph}$--$\rm n_{sph}$ relations defined by ETGs and LTGs (see Equations \ref{n_Msph_ETGs} and \ref{n_Msph_LTGs} for ETGs and LTGs, respectively). Galaxies excluded from our regressions, as discussed in Section \ref{Data}, are marked in magenta and cyan. Additionally, excluding the two extreme right LTGs (blue data-points) still yields consistent relation within $1\sigma$ uncertainty bound of the  $M_{\rm *, sph}$--$\rm n_{sph}$ relation for LTGs plotted here.}
\label{Msph-nmaj}
\end{center}
\end{figure*} 

We find two different relations in the $M_{\rm *,sph}$--$\rm n_{sph}$ diagram (Figure \ref{Msph-nmaj}) for the two morphological classes: ETGs and LTGs. Note that the  (galaxy absolute magnitude, $\rm \mathfrak{M}_{gal}$)--$\rm n$ relation for ETGs in  \citet{Young:Currie:1994}, \citet{Graham1996}, \citet{Jerjen2000}, \citet{Graham:Guzman:2003}, and \citet{FerrareseCote2006} pertains to the whole galaxy, not the spheroidal component of the ETG (unless it is an elliptical galaxy). The $\rm \mathfrak{M}_{sph}$--$\rm n$ relation in \citet{Andredakis1995}, \citet{Graham2001}, \citet{Khosroshahi:Wadadekar:Kembhavi:2000}, and \citet{Mollenhoff:Heidt:2001} pertains to the spheroid component of predominantly spiral galaxies. 

 The $M_{\rm *,sph}$--$\rm n_{sph,maj}$ relation that we derived for  ETGs can be expressed as
\begin{IEEEeqnarray}{rCl}
\label{n_Msph_ETGs}
\log(M_{\rm *,sph}/M_\odot) &=& (3.27\pm 0.25)\log\left(\rm n_{sph,maj}/3 \right) \nonumber \\
&& +\> (10.50\pm 0.06),
\end{IEEEeqnarray}
 with a total root mean square (rms) scatter of $\rm \Delta_{\rm rms|sph}= 0.46$ dex in the $\log(M_{\rm *,sph}) $-direction. The intrinsic scatter and correlation coefficients for Equation \ref{n_Msph_ETGs} and all other relations presented in this paper are provided in Tables \ref{fit parameters} and \ref{fit parameters2}. 
 As mentioned in Section \ref{Data}, we used the \textsc{BCES} bisector regression that treats the ordinate and abscissa symmetrically.  Additionally, 
 using the bisector line from the \textsc{MPFITEXY} regressions, we obtain the slope$=3.30 \pm 0.18$ and intercept$=10.50 \pm 0.04$,  which is closely consistent with the above relation obtained using the \textsc{BCES} regression. 
It should be noted that equation \ref{n_Msph_ETGs} is for spheroids, and is thus different from the (\textit{Galaxy} mass, $M_{\rm *,gal}$)--(galaxy S\'ersic index) relation for ETG sample containing disk galaxies.  
 
The bulges of LTGs follow a shallower relation which can be expressed as
 \begin{IEEEeqnarray}{rCl}
\label{n_Msph_LTGs}
\log(M_{\rm *,sph}/M_\odot) &=& (1.31\pm 0.22)\log\left(\rm n_{sph,maj}/3 \right) \nonumber \\
&& +\> (10.41\pm 0.07),
\end{IEEEeqnarray}
with  $\rm \Delta_{\rm rms|sph}= 0.32$ dex. 
The correlation of $M_{\rm *,sph}$ with the equivalent axis S\'ersic indices ($\rm n_{sph, eq}$) for ETGs and LTGs are consistent with the above Equations \ref{n_Msph_ETGs} and \ref{n_Msph_LTGs}, respectively, and are provided in Table \ref{fit parameters2}.
Equation \ref{n_Msph_LTGs} is also consistent with the relation obtained from the bisector \textsc{MPFITEXY} regression which provided the slope$=1.32 \pm 0.19$ and intercept$=10.41 \pm 0.06$ for LTGs. Similarly, for other correlations established in this paper, we have checked the best-fit lines using the \textsc{MPFITEXY} regression and these correlations with equivalent-axis bulge parameters are provided in the appendix Table \ref{fit parameters3}.
 
Our $M_{\rm *,sph}$--$\rm n$ relations for ETGs and LTGs support the dual sequences seen in the spheroid luminosity (absolute magnitude)--(S\'ersic index) diagram for ETGs and LTGs  by  \citet[][and references therein]{Savorgnan:n:2016}, which was based on a sub-sample of  our current sample.  Importantly, our greater sample size has enabled a reduced uncertainty on the slope and intercept of the relations. 

We also searched for substructures based on the other morphological information (core-S\'ersic vs S\'ersic galaxies, galaxies with a stellar disk versus galaxies without a stellar disk, and barred vs non-barred galaxies) and found no statistically significant division, except for a small difference between the best-fit lines for barred and non-barred galaxies (because the majority of our barred galaxies are LTGs)
 
Each of these relations implies that galaxies with greater spheroid stellar masses have higher spheroid S\'ersic indices \citep[][their figure 5]{Andredakis1995}, i.e., a higher central stellar light concentration. Moreover, the $M_{\rm *,sph}$--$\rm n_{sph}$ relations with different slopes for the two morphological types (ETGs and LTGs) imply two different progressions of spheroid mass with the central light concentration. This might be reflecting two different ways the stellar mass evolves in the bulges of ETGs and LTGs. Hence, these distinct relations should be helpful for simulations and semi-analytic models studying the formation and evolution of galaxies with different morphology.
We refrain from attempting a classical bulge versus pseudo-bulge classification. However, we note that no extra component for the (peanut shell)-shaped structure associated with a buckled bar \citep{Combes:Debbasch:1990, Athanassoula2015} is included in the galaxy decomposition because such features are effectively encapsulated by the B6 Fourier harmonic term \citep{Ciambur:2016:Profiler, Ciambur:Graham:2016,Ciambur2020} and the bar component of the decomposition. Inner discs are modelled as such.

\subsection{\textbf{The} \boldmath{$M_{\rm BH}-\rm n_{sph}$} \textbf{diagram}}
\label{3.2}
Obtaining the S\'ersic  index of a galactic spheroid is in some ways more straightforward than measuring its mass, or stellar velocity dispersion. This is because the S\'ersic index can be obtained from the decomposition of the galaxy light even if the image is not photometrically calibrated. Whereas, measuring the spheroid stellar mass requires decomposition of a flux-calibrated image, which further requires the distance to the galaxy and an appropriate stellar mass-to-light ratio. Similarly, the stellar velocity dispersion measurement requires reducing and analyzing telescope-time-expensive spectra of the central stars of the galaxy.

The correlation between black hole mass and S\'ersic index will, obviously, be beneficial for  estimating the black hole mass of a galaxy using the S\'ersic index of its spheroid (should it have one). \citet{GrahamErwin2003} were the first to establish a log-linear $M_{\rm BH}$--$\rm n_{sph}$ relation using a sample of 22 galaxies, which yielded $\rm \log M_{\rm BH} = (6.37 \pm 0.21) + (2.91 \pm 0.38) \log(\rm n_{sph})$. It had a comparable rms scatter of $\Delta_{\rm rms|BH}$=0.33 dex with the contemporary  $M_{\rm BH}$--$\sigma$ relation ($\Delta_{\rm rms|BH}$=0.31 dex) of the day. \citet{Graham:Driver:2007} subsequently advocated the log-quadratic relation  $\rm \log M_{\rm BH} = (7.98 \pm 0.09) + (3.70 \pm 0.46) \log(\rm n_{sph}/3) - (3.10 \pm 0.84) [\log \rm n_{sph}/3]^2 $, based on a sample of 27 galaxies. This resulted in a notably smaller intrinsic scatter (of just 0.18 dex) than that (0.31 dex) about their updated log-linear  relation $\rm \log M_{\rm BH} = (7.81 \pm 0.08) + (2.69 \pm 0.28) \log(\rm n_{sph}/3)$. In their log-quadratic $M_{\rm BH}$--$\rm n_{sph}$ relation,  galaxies with smaller S\'ersic indices resided on the steeper part of the curve, and galaxies with higher S\'ersic indices defined a shallower part of the curve. This might  have been an indication of  two different relations for low-$\rm n_{sph}$ and high-$\rm n_{sph}$ galaxies that they were not able to see because of a small sample.  

In consultation with the published literature, \citet{Savorgnan2013} doubled the sample size and derived the $M_{\rm BH}$--$n_{\rm sph}$ relations for S\'ersic and core-S\'ersic galaxies, however, the slopes of the two sub-samples were consistent within their $\pm 1\sigma$ uncertainty bound.  \citet{Savorgnan:n:2016} subsequently  used their own measurement of spheroid S\'ersic index  based on  multi-component decompositions, to establish a single log-linear $M_{\rm BH} \propto \rm n_{sph}^{(3.51 \pm 0.28)}$  relation, which was steeper than the relation reported by \citet{Graham:Driver:2007}. This is not surprising, as the slope from a single regression will vary arbitrarily according to the number of low- and high-n spheroids in one's sample.
This difference in the $M_{\rm BH}$--$\rm n_{sph}$ relation was also because \citet{Graham:Driver:2007} used the  forward (Y over X)  \textsc{FITEXY} regression routine from \citet{Tremaine:ngc4742:2002}, which minimized the  scatter in the quantity to be predicted, i.e., $M_{\rm BH}$, yielding a shallower slope for their $M_{\rm BH}$--$\rm n_{sph}$ relation. Though \citet{Graham:Driver:2007} did not calculate the bisector/symmetric-fit relation using the \textsc{FITEXY} routine, the \textsc{BCES} bisector regression over their dataset yielded a slope of $2.85\pm 0.40$ consistent with \citet{Savorgnan:n:2016}'s relation within the $\pm 1 \sigma$ uncertainty bound.
\citet{Savorgnan:n:2016} additionally explored the possibility of two different $M_{\rm BH}$--$n_{\rm sph}$ relations for ETGs and LTGs, however, due to just 17 LTGs in her sample, she could not find a statistically reliable best-fit line for the LTGs.  

Here, we reinvestigate the $M_{\rm BH}$--$\rm n_{sph}$ relation, roughly doubling the sample size of 64 from \citet{Savorgnan:n:2016}. Upon combining the latest $M_{\rm BH}$--$M_{\rm *,sph}$ relations for ETGs and LTGs from \citet{Sahu:2019:I} and \citet{Davis:2018:a} with our $M_{\rm *,sph}$--$\rm n_{sph}$ relations defined by ETGs and LTGs (Equations \ref{n_Msph_ETGs} and \ref{n_Msph_LTGs}), we expect $ M_{\rm BH} \propto \rm n_{sph}^{4.15 \pm 0.39}$ and $ M_{\rm BH} \propto \rm n_{sph}^{2.83 \pm 0.63}$ for ETGs and LTGs, respectively.

\begin{figure*}
\begin{center}
\includegraphics[clip=true,trim= 07mm 03mm 12mm 12mm,width=   1.0\textwidth]{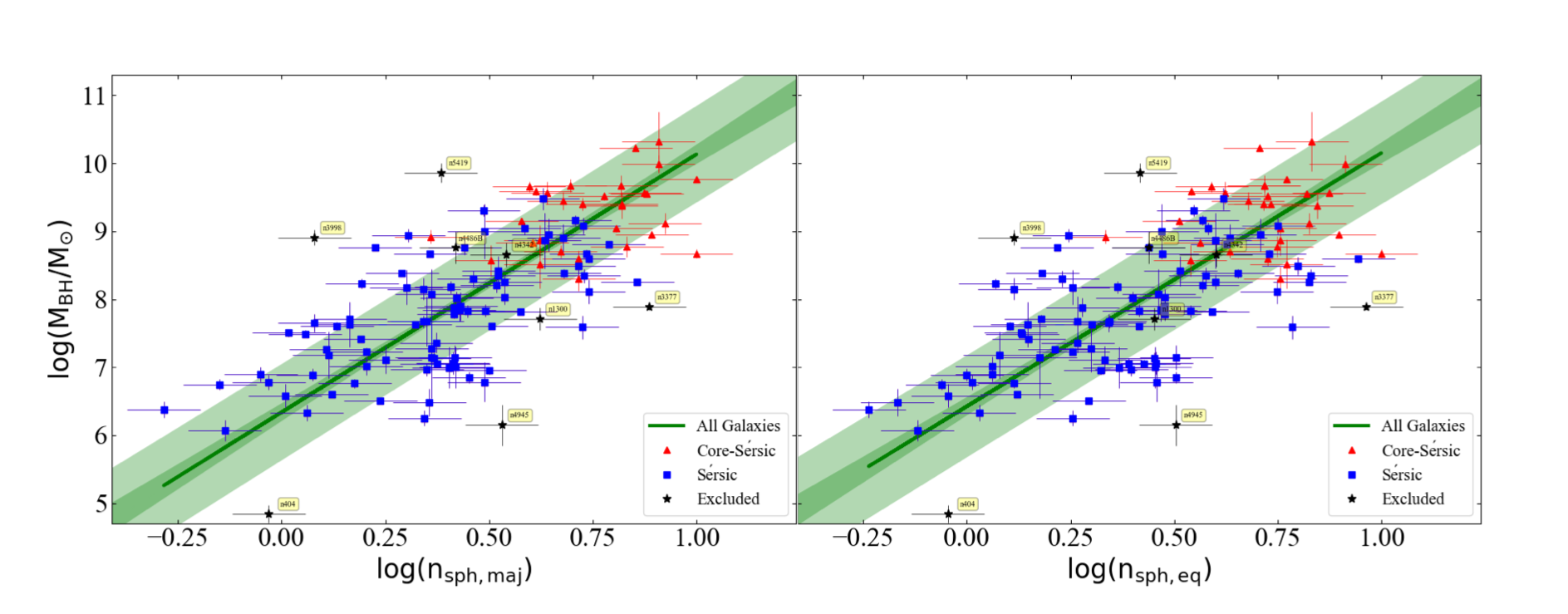}
\caption{Black hole mass versus major-axis (left panel) and equivalent-axis (right panel)  spheroid S\'ersic index.  S\'ersic and core-S\'ersic galaxies are shown in red and blue, respectively, and seem to follow the same single-regression $M_{\rm BH}$--$\rm n_{sph}$  relation.}
\label{Mbh-nmaj}
\end{center}
\end{figure*}

We started by performing a single symmetric regression between $M_{\rm BH}$ and $\rm n_{sph}$ for ETGs and LTGs combined (see  Figure \ref{Mbh-nmaj}), which gives 
\begin{IEEEeqnarray}{rCl}
\label{n_Mbh}
\log(M_{\rm BH}/M_\odot) &=& (3.79\pm 0.23)\log\left(\rm n_{sph,maj}/3 \right) \nonumber \\
&& +\> (8.15\pm 0.06),
\end{IEEEeqnarray}
between $M_{\rm BH}$ and $\rm n_{sph, maj}$  with a total rms scatter of $\Delta_{\rm rms| BH} = 0.69$ dex. Similarly, we obtained the single-regression relation between $M_{\rm BH}$ and $\rm n_{sph,eq}$, presented in Table \ref{fit parameters2}, which is closely consistent with the above  $M_{\rm BH}$--$\rm n_{sph,maj}$ relation.
Notably, this single-regression $M_{\rm BH}$--$\rm n_{sph,maj}$ relation  is consistent with the \citet{Savorgnan:n:2016} relation within her larger $\pm 1\sigma$ error bound of the slope and intercept. The asymmetric \textsc{BCES($M_{BH}|n$)} regression for our total sample yields $M_{\rm BH}$--$\rm n_{sph,maj}^{(3.15 \pm 0.22)}$, which is still consistent with the relation observed in \citet{Graham:Driver:2007}, again, within the $\pm 1 \sigma$ uncertainty limit of slopes. The intercept, however, has changed. This may partly be due to our use of majorly $\rm 3.6 \, \mu m$ data while \citet{Graham:Driver:2007} used R-band data\footnote{Many studies \citep[e.g.,][]{Kelvin:Driver:2012, Haussler:Bamford:2013, Kennedy:Bamford:2016}  have quantified the dependence of galaxy S\'ersic index on the wavelength band of image used.}.

\begin{figure*}
\begin{center}
\includegraphics[clip=true,trim= 07mm 03mm 12mm 12mm,width=   1.0\textwidth]{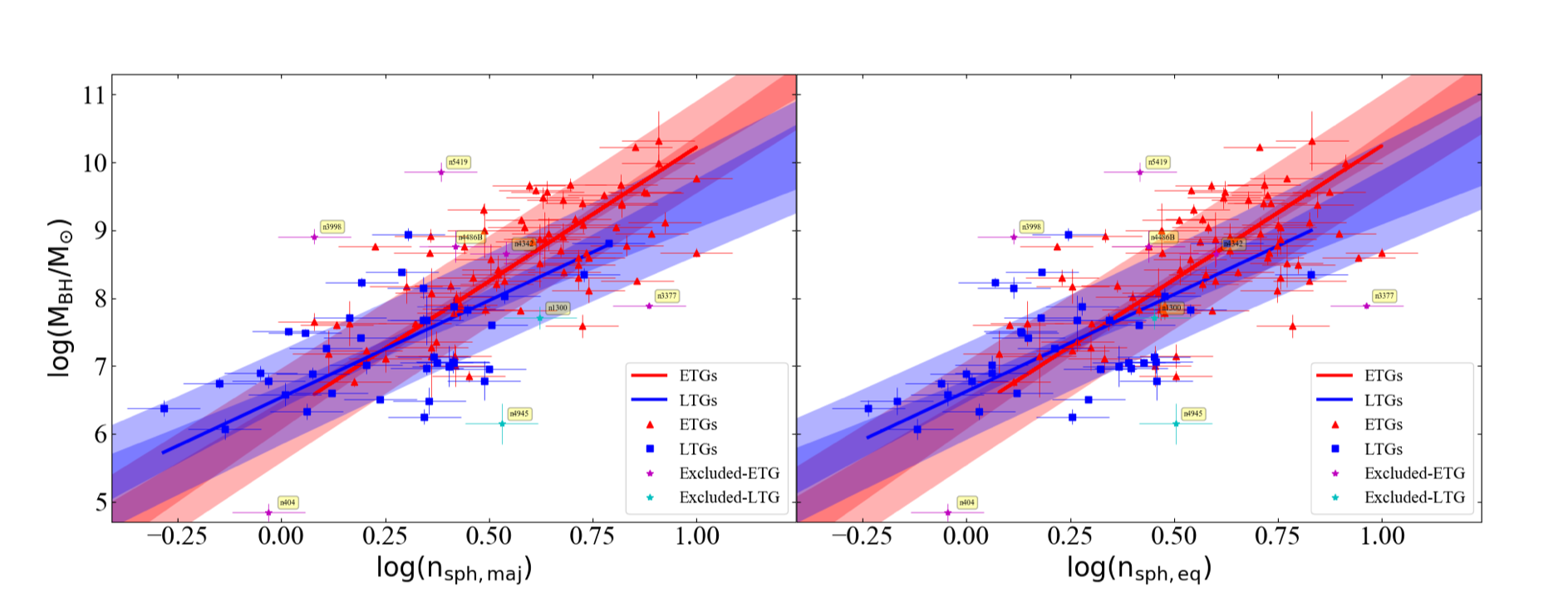}
\caption{Similar to Figure \ref{Mbh-nmaj}, but now showing the separate regressions for ETGs and LTGs as expressed in Equations \ref{n_Mbh_ETG} and \ref{n_Mbh_LTG}, resepectively. These relations are consistent with the predicted $M_{BH}$--$\rm n_{sph}$ relations obtained by combining the latest morphology-dependent $M_{BH}$--$M_{\rm *,sph}$ relations \citep{Davis:2018:a, Sahu:2019:I} with the $M_{\rm *,sph}$--$\rm n_{sph}$ relations from Figure \ref{Msph-nmaj}.}
\label{Mbh-nmaj2}
\end{center}
\end{figure*}

We further performed separate regressions for the  ETGs and LTGs. 
The symmetric $M_{\rm BH}$--$\rm n_{sph,maj}$ relation defined by ETGs can be expressed as
\begin{IEEEeqnarray}{rCl}
\label{n_Mbh_ETG}
\log(M_{\rm BH}/M_\odot) &=& (3.95\pm 0.34)\log\left(\rm n_{sph,maj}/3 \right) \nonumber \\
&& +\> (8.15\pm 0.08),
\end{IEEEeqnarray}
with $\Delta_{\rm rms| BH} = 0.65$ dex. The LTGs defined the shallower relation
\begin{IEEEeqnarray}{rCl}
\label{n_Mbh_LTG}
\log(M_{\rm BH}/M_\odot) &=& (2.85\pm 0.31)\log\left(\rm n_{sph,maj}/3 \right) \nonumber \\
&& +\> (7.90\pm 0.14),
\end{IEEEeqnarray}
with $\Delta_{\rm rms| BH} = 0.67$ dex. The $M_{\rm BH}$--$\rm n_{sph,maj}$ and  $M_{\rm BH}$--$\rm n_{sph,eq}$ relations obtained for ETGs versus LTGs are shown in the left- and right- hand panels of Figure \ref{Mbh-nmaj2}, respectively. The $M_{\rm BH}$--$\rm n_{sph,eq}$ relations for ETGs and LTGs are consistent with the above  $M_{\rm BH}$--$\rm n_{sph,maj}$ relations and are presented in Table \ref{fit parameters2}.
Importantly, the two relations for ETGs and LTGs in the $M_{\rm BH}$--$\rm n_{sph,maj}$ (and also in $M_{\rm BH}$--$\rm n_{sph,eq}$) diagram are consistent with the expected relations obtained after combining the $M_{\rm BH}$--$M_{*, sph}$ and  $M_{*, sph}$--$\rm n_{sph}$ relations (as mentioned before) for ETGs and LTGs within the $\pm 1 \sigma$ uncertainty bound. 

We also performed multiple double regressions by dividing our sample into S\'ersic versus core-S\'ersic galaxies, galaxies with a disk (ES-, S0-, Sp-Types) versus galaxies without a disk (E-Type), and barred versus non-barred galaxies. In the former two cases, we did not find statistically different relations. Whereas, we see two slightly different $M_{\rm BH}$--$\rm n_{sph}$ lines for barred and non-barred galaxies because most of our  LTGs are barred while most of our ETGs are non-barred. Moreover, the difference between the two relations followed by ETGs and LTGs is more prominent and consistent with the expected relations; hence, we conclude that the substructure in the $M_{\rm BH}$--$\rm n_{sph}$ diagram is due to ETG versus LTG categorization. Notably higher scatters around the $M_{\rm BH}$--$\rm n_{sph}$ relations depicted  in Figure \ref{Mbh-nmaj2} obstruct the visibility of distinct relations for ETGs and LTGs, even though our statistical analysis suggests different relations. We reckon that, in future, a bigger data set will enable visibly distinct  $M_{\rm BH}$--$n_{\rm sph}$ relations defined by ETGs and LTGs.
For a comparison with the barred versus non-barred case, we also provide  the  $M_{\rm BH}$--$\rm n_{sph,maj}$ (and  $M_{\rm BH}$--$\rm n_{sph,eq}$) relations obtained for the barred and non-barred galaxies along with  the relations for ETGs and LTGs in Tables \ref{fit parameters} and \ref{fit parameters2}.

\subsubsection{\textbf{The} \boldmath{$M_{\rm BH}$--$\rm Concentration$} \textbf{diagram}}
\label{3.2.1}
We also analyzed the relation between black hole mass and the light concentration of spheroids.  
\citet{Trujillo:Graham:Caon:2001} quantified a central concentration index, for the light profile captured by a S\'ersic function, as  \enquote{a flux ratio} which can be expressed as $\rm C (\alpha) = \gamma (2n, b_n  \alpha^{1/n})/\gamma (2n, b_n)$.  Where, $\alpha $ is equal to  $\rm r/R_e$, and $0< \alpha <1$.  For a particular $\alpha$, a higher value of $\rm C(\alpha)$ represents a spheroid or an elliptical galaxy with a greater central light or mass concentration. 

\begin{figure}
\begin{center}
\includegraphics[clip=true,trim= 10mm 05mm 20mm 20mm,width=   1.0\columnwidth]{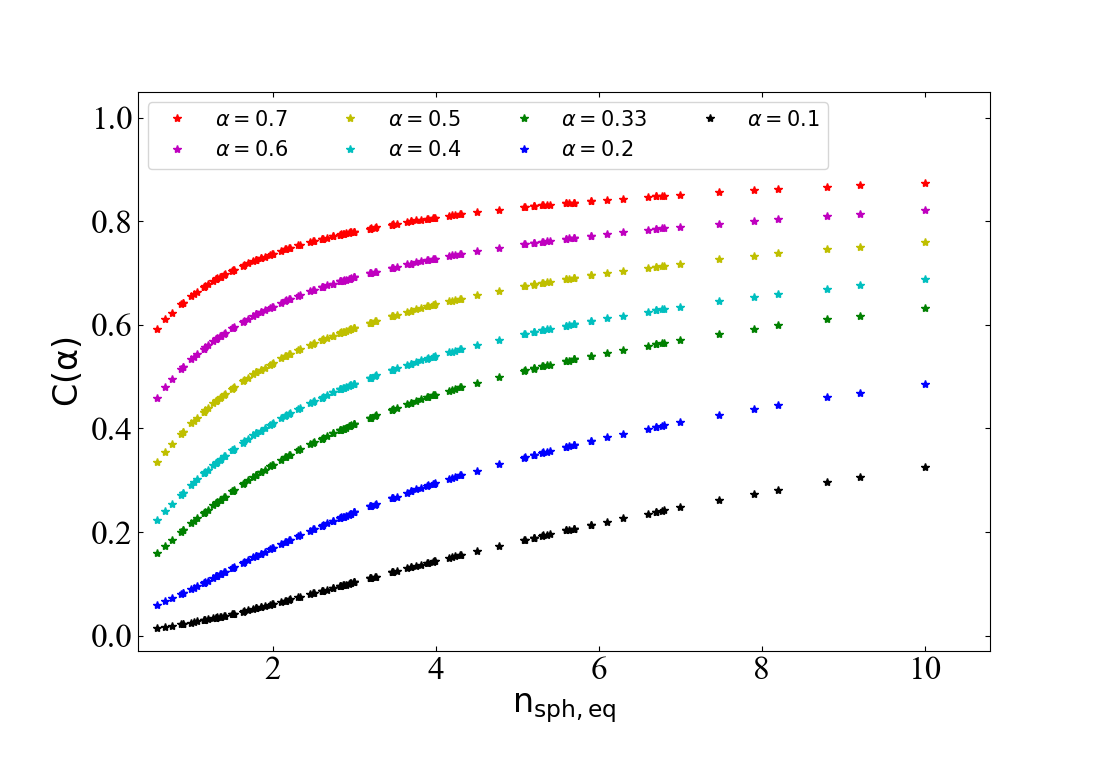}
\caption{Central light concentration index plotted against equivalent axis S\'ersic index for a range of $\alpha$ (fraction of effective half-light radius), representing the monotonicity between the concentration index and the S\'ersic index. This plot also shows that for a high  value of $\alpha$ ($\gtrsim$0.5), the range of $\rm C(\alpha)$ values is very small such that the increment in the $\rm C(\alpha)$ with increasing n becomes minimal for $\rm n \gtrsim 2$.}
\label{C-n}
\end{center}
\end{figure}

To calculate the concentration index for our spheroids, we use the equivalent axis S\'ersic index and the exact value of $\rm b_n$ obtained using the equation $\Gamma(2n) = 2\gamma(2n, b_n)$. In Figure \ref{C-n}, we have plotted $\rm C(\alpha)$ for our spheroids, for a range of $\alpha$ values, against their equivalent axis S\'ersic indices, revealing how both quantities are related monotonically, as already seen in \citet{Trujillo:Graham:Caon:2001}. 

\citet{GrahamErwinCaon2001} explored a range of values of $\alpha$ and found that $\alpha =1/3$ produces a minimum scatter in the vertical direction in the $M_{\rm BH}$--$\rm C(\alpha)$ diagram. Moreover, for $\alpha > 0.5$ the range of concentration index values is so small that it becomes indistinguishable for different profile shapes  (i.e., n), which is evident in our Figure \ref{C-n}, while low values of $\alpha (< 0.2)$ are not so practical, especially for high redshift galaxies, as they require high spatial resolution   \citet{GrahamTrujilloCaon2001}. 
Therefore, in our investigation of the $M_{\rm BH}$--$\rm C(\alpha)$ relation, we use $\rm C(\alpha)$ at $\alpha =1/3$ for our spheroids. The uncertainty in $\rm C(1/3)$ is calculated via error propagation based on a $20\%$ uncertainty in the S\'ersic index.

\begin{figure}
\begin{center}
\includegraphics[clip=true,trim= 10mm 05mm 20mm 20mm,width= 1.0\columnwidth]{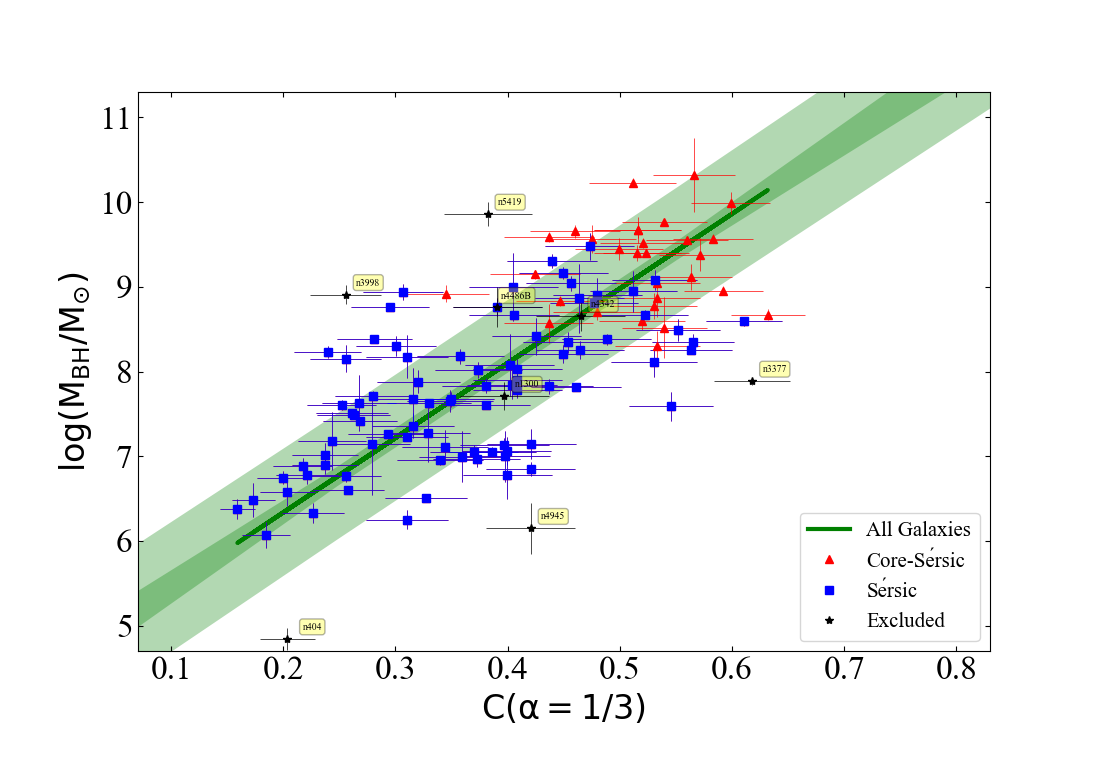}
\caption{Black hole mass versus the spheroid's central concentration index calculated using the equivalent-axis  S\'ersic index.}
\label{Mbh-CI}
\end{center}
\end{figure}

The correlation we obtained upon performing a symmetric regression between $M_{\rm BH}$ and $\rm C(1/3)$  for the total (ETGs+LTGs) sample can be expressed as, 
\begin{IEEEeqnarray}{rCl}
\label{CI_Mbh}
\log(M_{\rm BH}/M_\odot) &=& (8.81\pm 0.53)\log\left[\rm C(1/3) /0.4 \right] \nonumber \\
&& +\> (8.10\pm 0.07).
\end{IEEEeqnarray}
with $\Delta_{\rm rms| BH} = 0.73$ dex in $M_{\rm BH}$-direction. This is represented in Figure \ref{Mbh-CI}. This relation is steeper than the relation $M_{\rm BH} \propto \rm C(1/3)^{(6.81\pm 0.95)}$ reported by \citet{GrahamErwinCaon2001} which was based on a set of only 21 galaxies.

\begin{figure*}
\begin{center}
\includegraphics[clip=true,trim= 07mm 03mm 12mm 12mm,width=   1.0\textwidth]{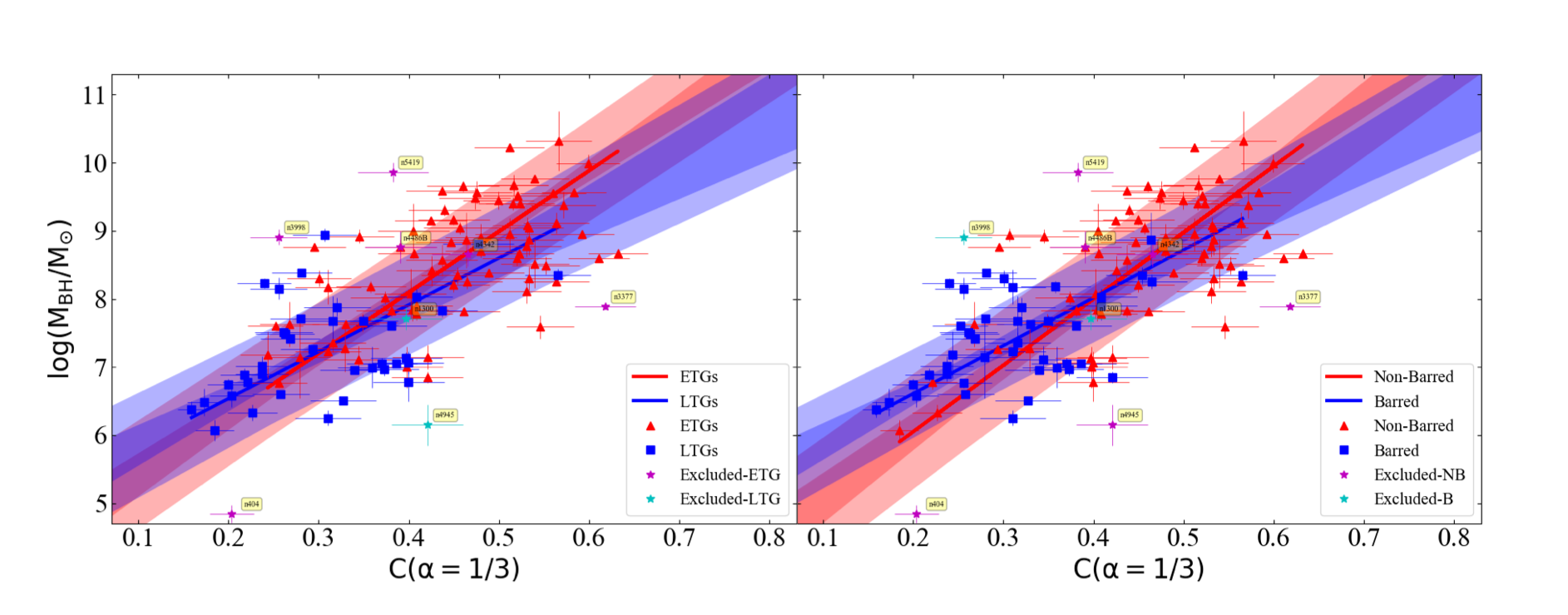}
\caption{Similar to Figure \ref{Mbh-CI}, but now showing the best-fit lines obtained for  ETGs and LTGs (left panel) plus barred and non-barred galaxies (right panel). The  different  lines obtained for barred and non-barred galaxies (right panel) is a consequence of most of our  barred galaxies being LTGs.}
\label{Mbh-CI2}
\end{center}
\end{figure*}

Here, again, we looked for substructures due to S\'ersic versus core-S\'ersic galaxies, galaxies with a disk versus galaxies without a disk, barred versus non-barred galaxies, and ETGs versus LTGs. We find two slightly different relations only for the latter two cases, similar to the $M_{\rm BH}$--$\rm n_{sph}$ diagram, which is represented in Figure \ref{Mbh-CI2}.  Again, the substructure in the  $M_{\rm BH}$--$\rm C(\alpha)$ diagram due to barred and non-barred galaxies is likely due to most of the LTGs being barred, while the dominant substructuring is due to the ETG and LTG morphology. The parameters of the  $M_{\rm BH}$--$\rm C(1/3)$ relations defined by ETGs and LTGs are provided in Table \ref{fit parameters2}. The best-fit lines obtained for the barred and non-barred galaxies are also provided in Table \ref{fit parameters2} for comparison.

\subsection{\textbf{The} \boldmath{$M_{\rm \rm *,sph}-\rm R_{e,sph}$}\textbf{diagram}}
\label{3.3}

There is a long history of studies  which have worked on the galaxy size--luminosity ($\rm L_{gal}$--$\rm R_{e,gal}$) relation for ETGs and found it to be curved  \citep[see][for a review]{Graham:Re:2019}.
Here we explore the $M_{\rm *,sph}$--$\rm R_{e,sph}$ diagram for the spheroids of ETGs and LTGs in our sample, for whom $\rm R_{e,sph}$ values were obtained from a careful image analysis.

\begin{figure*}
\begin{center}
\includegraphics[clip=true,trim= 07mm 03mm 12mm 12mm,width=   1.0\textwidth]{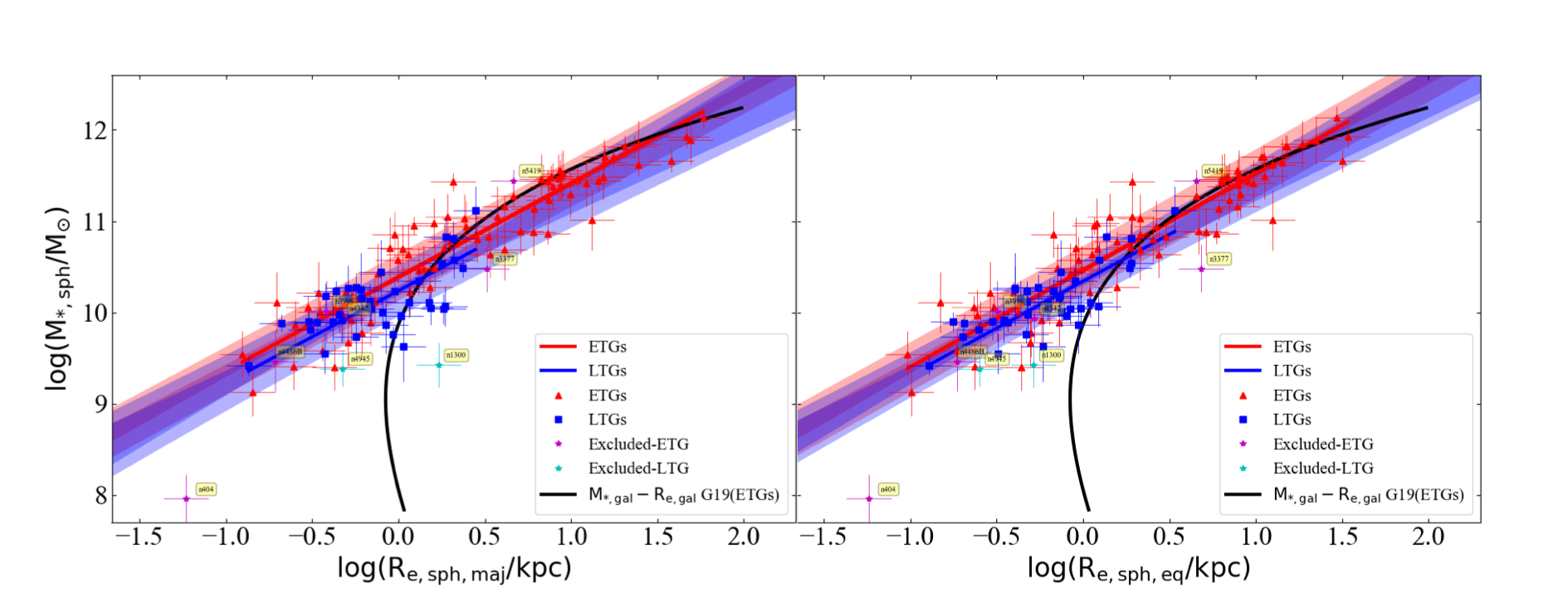}
\caption{Spheroid stellar mass versus major-axis (left panel) and equivalent-axis (right panel) effective half-light radius of the spheroid. Both  panels reveal that the spheroids of ETGs and LTGs follow closely consistent relations suggesting that a single $M_{\rm *,sph}$--$\rm R_{e,sph}$ relation (Equation \ref{Re_Msph}) for all galaxy types is sufficient for the current data-set. The black  curve is the $M_{\rm *,gal}$--$\rm R_{e,gal}$ relation for ETGs taken from \citet[][their Figure 18]{Graham:Re:2019} abbreviated as \enquote{G19}.}
\label{Msph-Re}
\end{center}
\end{figure*}

Upon performing two different regressions for our ETGs and LTGs, we find a tight correlation between $M_{\rm *, sph}$ and $\rm R_{e,sph}$ (see Figure \ref{Msph-Re}) for both cases with remarkably smaller scatter ($\Delta_{\rm rms | sph}  \sim$ 0.26 dex) in the $M_{\rm *, sph}$-direction than the $M_{\rm \rm *,sph}$--$\rm n_{sph}$ relations\footnote{The rms scatters in the horizontal direction for Equations \ref{n_Msph_ETGs}, \ref{n_Msph_LTGs}, and \ref{Re_Msph} are $\rm \Delta_{\rm rms|n}=$0.14 dex, $\rm \Delta_{\rm rms|n}=$0.24 dex, and $\rm \Delta_{\rm rms|Re}=$0.25 dex, respectively.} (Equations \ref{n_Msph_ETGs} and \ref{n_Msph_LTGs} with $\rm \Delta_{\rm rms|sph}=$ 0.46 and 0.32 dex). The left- and right-hand panels in Figure \ref{Msph-Re} show the major-axis and equivalent-axis effective half-light radii ($\rm R_{e,sph, maj}$ and $\rm R_{e,sph, eq}$), respectively, on the horizontal-axes. The parameters for the $M_{\rm *, sph}$--$\rm R_{e,sph}$ relations for both ETGs and LTGs are provided in Tables \ref{fit parameters} (major-axis) and \ref{fit parameters2} (equivalent-axis). 

\begin{figure*}
\begin{center}
\includegraphics[clip=true,trim= 07mm 03mm 12mm 12mm,width=   1.0\textwidth]{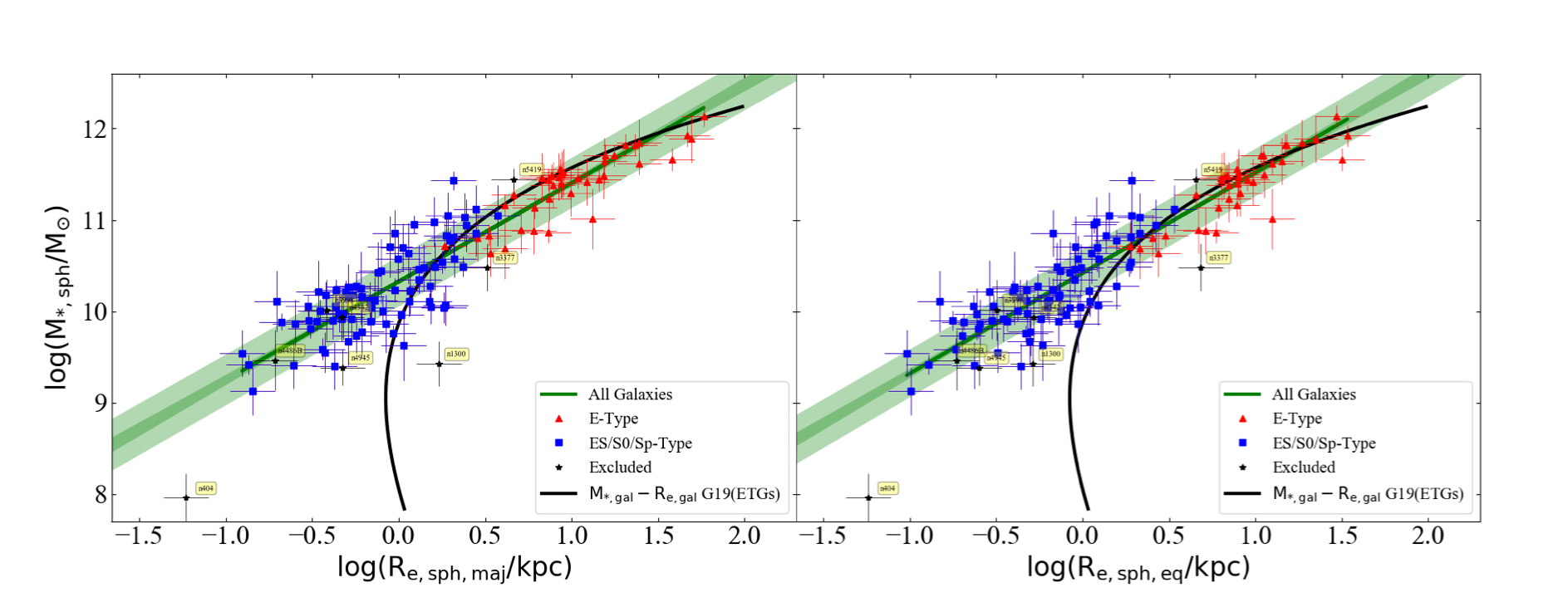}
\caption{Similar to Figure \ref{Msph-Re}, but now showing the single-regression $M_{\rm \rm *,sph}$--$\rm R_{e,sph}$ relation defined by the total (ETG+LTG) sample.}
\label{Msph-Re_single}
\end{center}
\end{figure*}

The best-fit $M_{\rm *, sph}$--$\rm R_{e,sph}$ lines for both ETGs and LTGs are log-linear and very close,  such that their $\pm 1 \sigma$ scatter region (shaded red and blue area in Figure \ref{Msph-Re}) almost overlap with each other. 
Therefore, we further perform a single symmetric regression  for our total (ETG+LTG) sample, obtaining
\begin{IEEEeqnarray}{rCl}
\label{Re_Msph}
\log(M_{\rm *,sph}/M_\odot) &=& (1.08\pm 0.04)\log\left(\rm R_{e,sph, maj}/kpc \right) \nonumber \\
&& +\> (10.32\pm 0.03),
\end{IEEEeqnarray}
with $\Delta_{\rm rms|sph} = 0.27$ dex. This single-regression is represented in Figure \ref{Msph-Re_single}, where the left-hand and right-hand panels show the $M_{\rm *, sph}$--$\rm R_{e,sph,maj}$ and $M_{\rm *, sph}$--$\rm R_{e,sph,eq}$ relations, respectively.
The parameters for the single-regression $M_{\rm *, sph}$--$\rm R_{e,sph,eq}$ relation can be found in Table \ref{fit parameters2}, which has consistent slope with the  above $M_{\rm *, sph}$--$\rm R_{e,sph,maj}$ relation. 

Our total (ETG+LTG) sample also includes some alleged  pseudo-bulges, 
marked in Table-1 of \citet{Sahu:2019:II} along with their source of identification, suggesting that the above single log-linear $M_{\rm *, sph}$--$\rm R_{e,sph}$ relation applies for both alleged pseudo-bulges and the normal/classical bulges. 

For a comparison, we have plotted the $M_{\rm *,gal}$--$\rm R_{e,gal}$ curve for ETGs from \citet[][their Figure 18]{Graham:Re:2019} in our Figures \ref{Msph-Re} and \ref{Msph-Re_single}. The shallower part of this curve, at the high mass (and size) end, seems to match well with our near-linear $M_{\rm *, sph}$--$\rm R_{e,sph}$ relation for bulges, however, the  $\rm R_{e,sph}$  of our spheroids becomes smaller than their $\rm R_{e,gal}$ at $\rm \log(M_{\rm *,sph}) \lesssim \, 10.5 \, dex$ (or $\rm R_e \lesssim  2 \, kpc$) due to the presence of disks enabling bigger $\rm R_{e,gal}$ for their ETGs\footnote{This is also partly intuitive because, for a given stellar density,  a 2D disk (or a galaxy with a dominant disk) having the same total stellar mass as a 3D spheroidal distribution of stars will extend to a larger radii.}. 
We do not obtain a curved $M_{\rm *, sph}$--$\rm R_{e,sph}$ relation, possibly, because our sample does not include many dwarf/low-mass  ETGs or late-type spiral (Sc-, Sd-types) galaxies. 

The bend-point of the curved $\rm L_{gal}$--$\rm R_{e,gal}$ relation for ETGs has been of  past interest, because many studies have claimed that this bend-point is the point of distinction between dwarf elliptical (dE) and classical spheroids or (normal) elliptical galaxies \citep{Sersic:1968:2, Kormendy:Fisher:2009, FisherDrory2010, Fisher:2016}. Different  physical formation processes have been invoked for these alleged disjoint classes of galaxies  \citep[e.g.,][]{Tolstoy2009, Kormendy:Bender:2012, Somerville:Dave:2015}. Providing a detailed investigation of this curved relation, \citet[][their figure 4]{Graham:Re:2019} present a (galaxy luminosity)--$\rm R_z$ diagram,  where $\rm R_z$  represents the radius of the projected galaxy image enclosing z\% of the total light, for z varying from 2\% to 97\%,  including $\rm R_e$ for which z=50\%.  \citet{Graham:Re:2019} find that all the $\rm L_{B, gal}$--$\rm R_{z_{gal}}$ relations are curved but the location (the absolute magnitude) of the bend-point of each curve changes with z, revealing that the bend-point in the $\rm L$--$\rm R_{e \, (or \, z=50 \%)}$ relation has been used to artificially divide galaxies at a random magnitude based on the random percentage of light used to measure galaxy sizes. 

\begin{figure*}
\begin{center}
\includegraphics[clip=true,trim= 06mm 03mm 10mm 10mm,width=   1.0\textwidth]{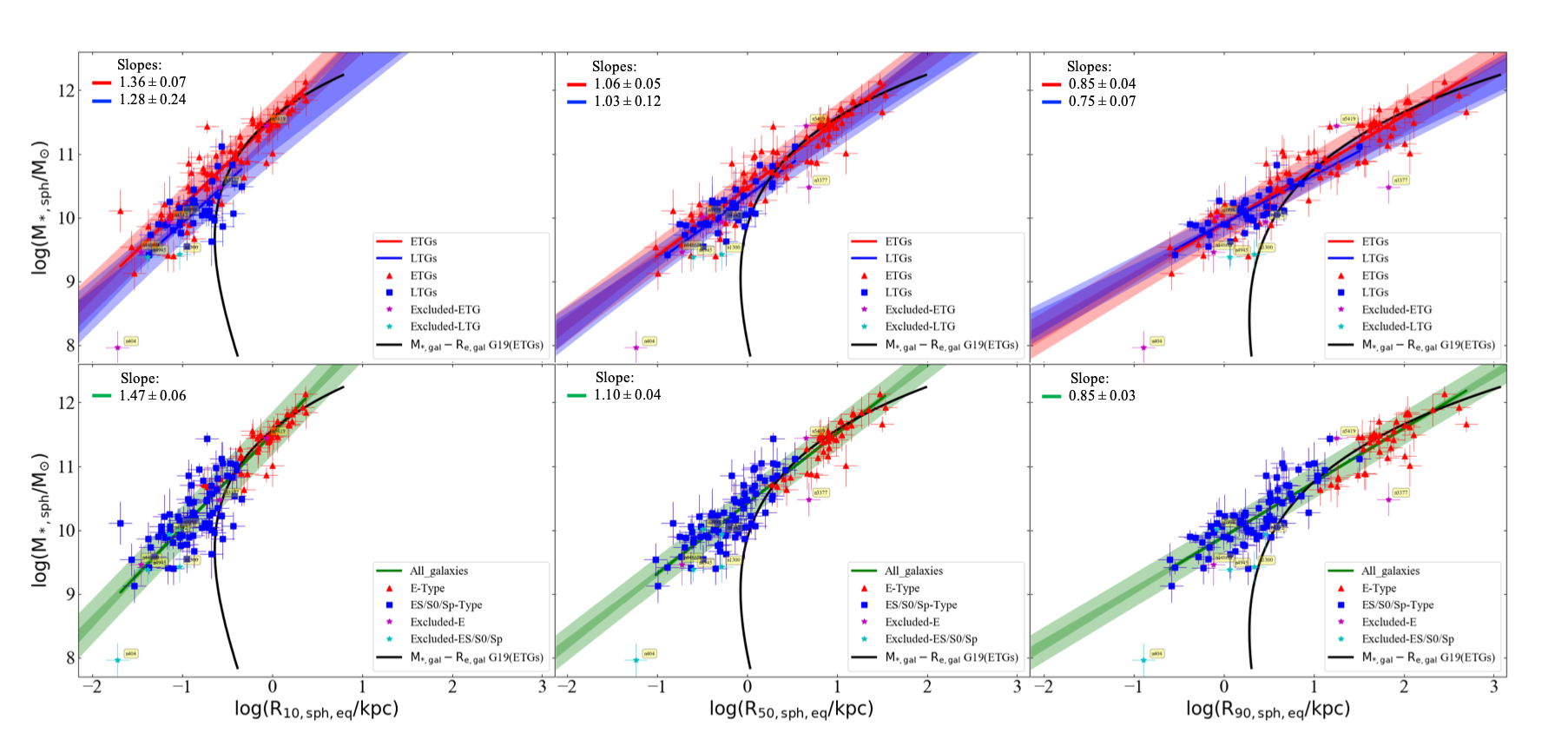}
\caption{Similar to Figure \ref{Msph-Re} and \ref{Msph-Re_single}, but now showing  $\rm R_{10,sph,eq}$,  $\rm R_{50,sph,eq}$ (or  $\rm R_{e,sph,eq}$), and $\rm R_{90,sph,eq}$ on the horizontal axis  in the left, middle, and right panels, respectively. Black curves are the $M_{\rm *,gal}$--$\rm R_{z,gal}$ curves from \citet{Graham:Re:2019} for the corresponding percentage (\enquote{$\rm z\%$}) of  enclosed light.
The top panels demonstrate that in all three cases ETGs and LTGs define consistent  relations between $M_{\rm *, sph}$ and $\rm R_{z,sph}$ for the range of our sample,  suggesting a single $M_{\rm *,sph}$--$R_{\rm z,sph}$ relation for the total (ETG+LTG) sample, presented in the bottom panels for each case. Importantly, the $M_{*, sph}$--$\rm R_{z,sph}$ relations become  shallower with increasing z. All the  parameters for the $M_{*, sph}$--$\rm R_{z,sph}$ relations are provided in Table \ref{fit parameters2}.}
\label{combo_Msph}
\end{center}
\end{figure*}

Following \citet{Graham:Re:2019}, using their  Equation 22, we also calculated the radii containing $\rm z=10\%$ and $\rm z=90 \%$ of the spheroid's light, i.e., $\rm R_{10,sph}$ and $\rm R_{90,sph}$, respectively. Figure \ref{combo_Msph} demonstrates how the spheroid stellar mass correlates with the equivalent axis radii $\rm R_{10,sph,eq}$,  $\rm R_{50,sph,eq}$ (or  $\rm R_{e,sph,eq}$), and $\rm R_{90,sph,eq}$, in the left, middle, and right panels, respectively. In all three cases, we find that ETGs and LTGs follow consistent relations suggesting a single $M_{*, sph}$--$\rm R_{z,sph}$ relation in each panel, however, the slope (and intercepts) of the relations change gradually with z. For comparison, we also show the $M_{\rm *,gal}$--$\rm R_{z,gal}$ curves from \citet{Graham:Re:2019}, which seem to agree well with the elliptical galaxies at the high mass end of our $M_{\rm *, sph}$--$\rm R_{z,sph}$ relations. Whereas for galaxies with a disk (i.e., ES-, S0-, Sp-types), the radius containing $z\%$ of the spheroid's light ($\rm R_{z,sph}$) is  smaller than the radius containing $z\%$ of whole galaxy's light ($\rm R_{z,gal}$). The parameters for the $M_{\rm *, sph}$--$\rm R_{10,sph,eq}$ and  $M_{\rm *, sph}$--$\rm R_{90,sph,eq}$ relations are also provided in Table \ref{fit parameters2}. Though for the range of our data-set we observe a (log)-linear relation between $M_{\rm *, sph}$ and $\rm R_{z,sph}$, addition of galaxies at the low-mass and small size end might reveal a curved $M_{\rm *, sph}$--$\rm R_{z,sph}$  relation similar to the $M_{\rm *,gal}$--$\rm R_{z,gal}$ curve for ETGs.

\subsection{\textbf{The} \boldmath{$M_{\rm BH}$--$\rm R_{e,sph}$} \textbf{diagram}} 
\label{3.4}
Combining the $M_{\rm BH}$--$M_{\rm *,sph}$ relations defined by ETGs and LTGs, from \citet{Sahu:2019:I} and \citet{Davis:2018:a}, with the single-regression $M_{\rm *,sph}$--$\rm R_{e,sph, maj}$ relation (Equation \ref{Re_Msph}) followed by our combined sample of ETGs and LTGs, we expect $M_{\rm BH} \propto \rm R_{e,sph, maj}^{1.37 \pm 0.09}$ and $M_{\rm BH} \propto \rm R_{e,sph, maj}^{2.33 \pm 0.36}$  for ETGs and LTGs, respectively.

We first used a single regression for the total (ETG+LTG) sample, which yielded a good relation, provided in Tables \ref{fit parameters} and \ref{fit parameters2} for the major- and equivalent-axis $\rm R_{e,sph}$, respectively,  but with a higher scatter than the soon to be revealed separate relations for ETGs and LTGs. Moreover, it is inconsistent with the above prediction of the two relations in the  $M_{\rm BH}$--$\rm R_{e,sph}$ diagram.
 
\begin{figure*}
\begin{center}
\includegraphics[clip=true,trim= 07mm 03mm 12mm 12mm,width=   1.0\textwidth]{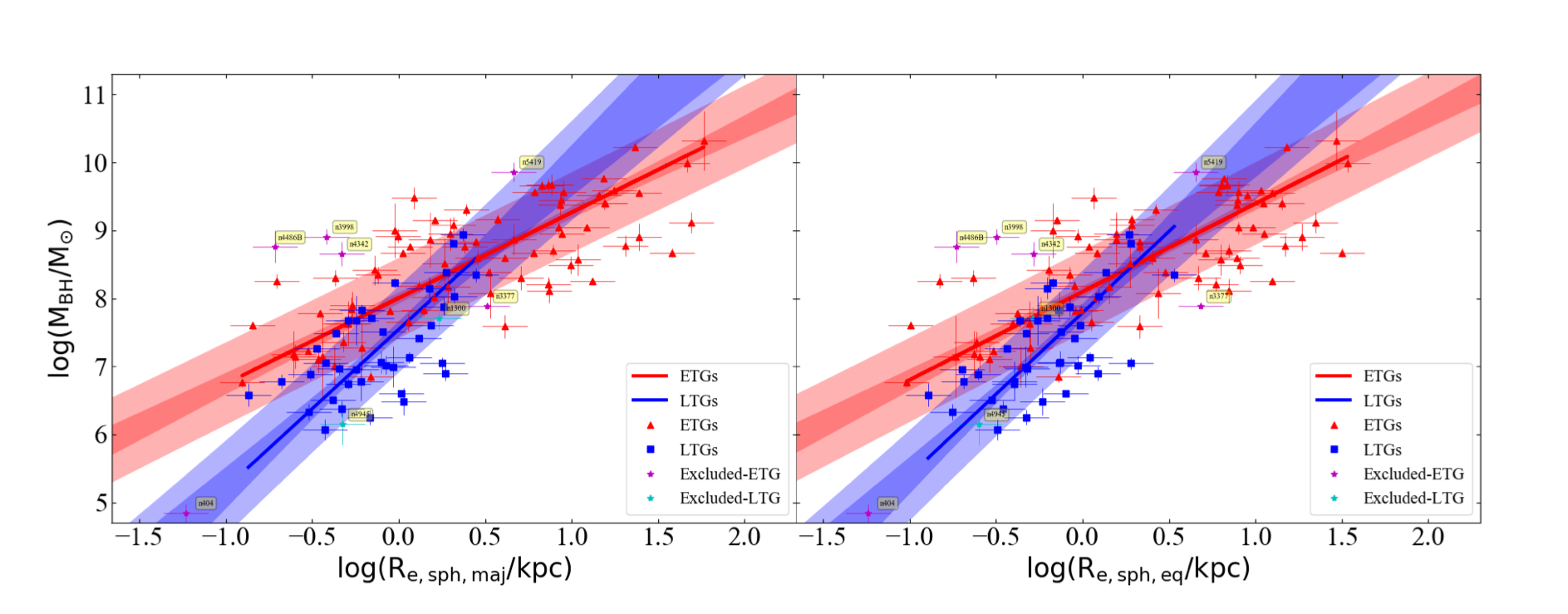}
\caption{Black hole mass versus major-axis (left panel) and equivalent-axis (right panel) effective half-light radius of the spheroid. Both panels reveal that ETGs and LTGs follow two different $M_{\rm BH}$--$\rm R_{e,sph}$  relations (Equations \ref{Re_Mbh_ETG} and \ref{Re_Mbh_LTG}).}
\label{MBH-Re}
\end{center}
\end{figure*}

Upon performing separate regressions for ETGs and LTGs in the $M_{\rm BH}$--$\rm R_{e,sph}$ diagram, we do find two different  relations for the two morphological classes. These relations are presented in Figure \ref{MBH-Re} with the  left-hand and right-hand panels displaying $\rm R_{e,sph, maj}$ and $\rm R_{e,sph, eq}$, respectively.
The relation defined by all ETGs can be expressed as,
\begin{IEEEeqnarray}{rCl}
\label{Re_Mbh_ETG}
\log(M_{BH}/M_\odot) &=& (1.26\pm 0.08)\log\left(\rm R_{e,sph, maj}/kpc \right) \nonumber \\
&& +\> (8.00\pm 0.07),
\end{IEEEeqnarray}
with $\Delta_{\rm rms|BH} = 0.58$ dex, while LTGs define the relation
\begin{IEEEeqnarray}{rCl}
\label{Re_Mbh_LTG}
\log(M_{BH}/M_\odot) &=& (2.33\pm 0.31)\log\left(\rm R_{e,sph, maj}/kpc \right) \nonumber \\
&& +\> (7.54\pm 0.10),
\end{IEEEeqnarray}
with $\Delta_{\rm rms|BH} = 0.62$ dex. The slope of the $M_{\rm BH}$--$\rm R_{e,sph,eq}$ relations for  ETGs and LTGs are consistent with the corresponding $M_{\rm BH}$--$\rm R_{e,sph, maj}$ relations, and their fit parameters are provided in Table \ref{fit parameters2}. These two relations (Equations \ref{Re_Mbh_ETG} \& \ref{Re_Mbh_LTG}) for ETGs and LTGs are in agreement with the expected $M_{\rm BH}$--$\rm R_{e,sph}$ relations mentioned at the beginning of this sub-section.
Additionally, we note that our $M_{\rm BH}$--$\rm R_{e,sph, maj}$ relation for ETGs is also consistent with the relation obtained by \citet{Nicola:2019}, based on an ETG-dominated sample. 

\begin{figure*}
\begin{center}
\includegraphics[clip=true,trim= 07mm 03mm 12mm 12mm,width=   1.0\textwidth]{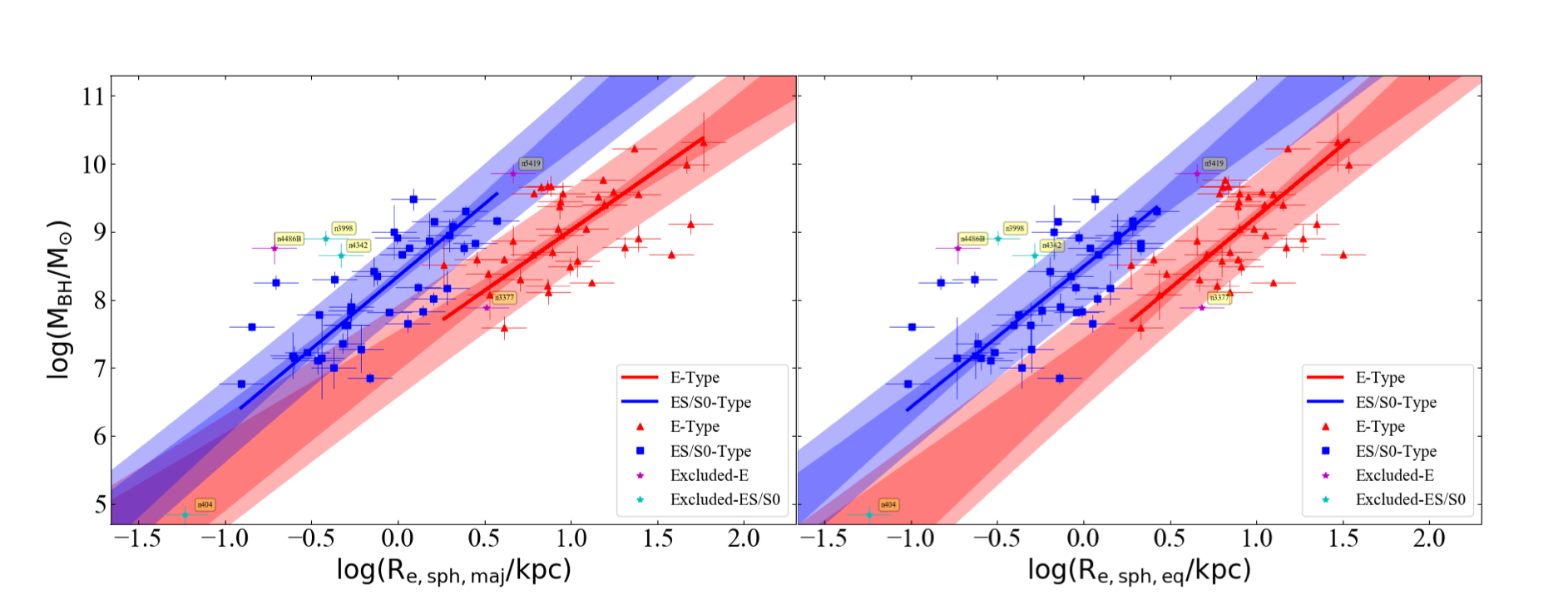}
\caption{Similar to Figure \ref{MBH-Re}, but now showing only the ETGs with a disk (ES and S0-types) and ETGs without a disk (E-type), which define two almost parallel $M_{\rm BH}$--$\rm R_{e,sph}$ relations (listed in Tables \ref{fit parameters} and \ref{fit parameters2}) with an offset of $\sim $1 dex in the vertical direction. This explains the related offset in the $M_{\rm BH}$--$ M_{\rm *,sph}$ diagram \citep{Sahu:2019:I}. }
\label{MBH-Re_disk}
\end{center}
\end{figure*}

\begin{figure*}
\begin{center}
\includegraphics[clip=true,trim= 07mm 03mm 12mm 12mm,width=   1.0\textwidth]{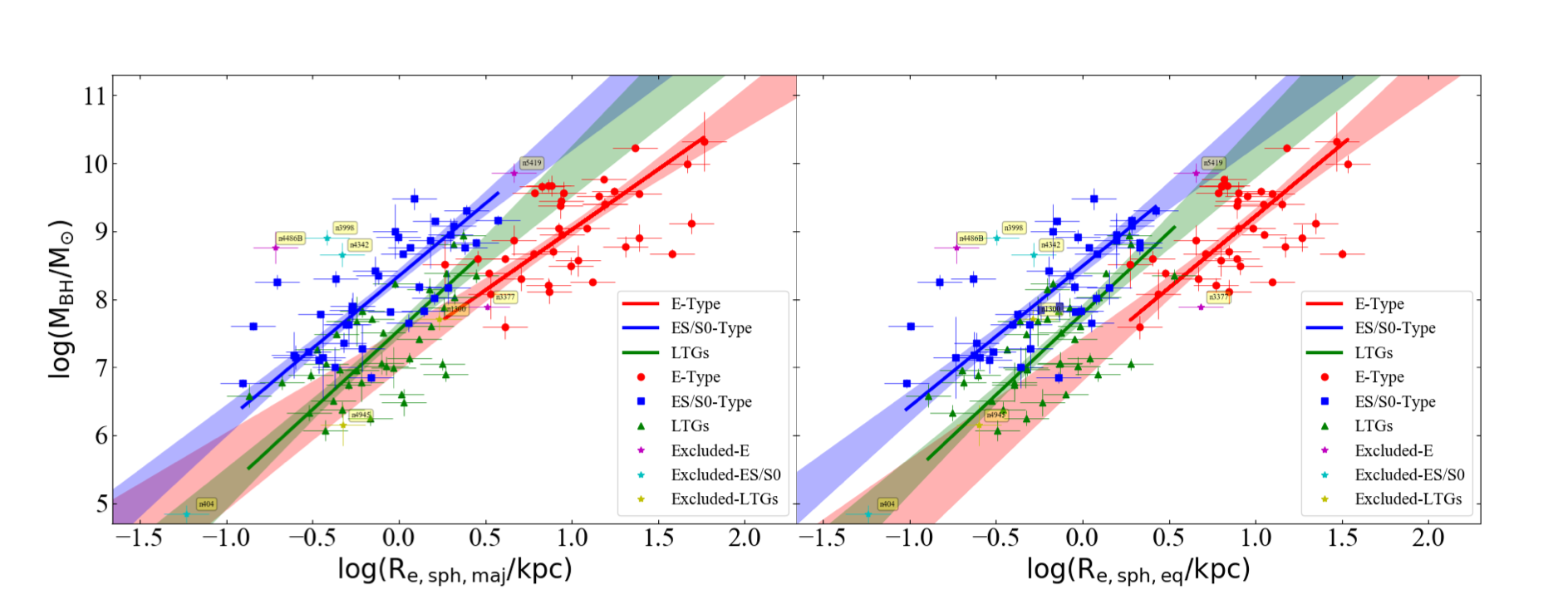}
\caption{Similar to Figure \ref{MBH-Re_disk}, but now also showing the $M_{\rm BH}$--$\rm R_{e,sph}$ relation defined by LTGs in the same diagram. Just for clarity, we are not showing the (light-shaded) $1 \sigma$ scatter regions, which are visible in Figures \ref{MBH-Re} and \ref{MBH-Re_disk}.}
\label{MBH-Re_triple}
\end{center}
\end{figure*}

Each of our non-linear, but log-linear, $M_{\rm BH}$--$\rm R_{e,sph}$ relations reveal that galaxies with more massive black holes tend to have a larger (bulge) half-light radii. However, the two different slopes of the $M_{\rm BH}$--$\rm R_{e,sph}$ relations for ETGs and LTGs suggest that the process of evolution between black hole mass and spheroid size ($R_{e,sph}$), which further relates to the spheroid stellar mass, tends to be different for these different morphological types. 
 This also supports our morphology-dependent $M_{\rm BH}$--$M_{\rm \rm *,sph}$,  and $M_{\rm BH}$--$M_{\rm \rm *,gal}$ relations \citep{Davis:2018:b, Sahu:2019:I}, where ETGs and LTGs are found to follow two different relations. The total rms scatter about the $M_{\rm BH}$--$\rm R_{e,sph}$ relation is smaller than the total rms scatter about the  $M_{\rm BH}$--$M_{\rm *,sph}$ relation for LTGs (cf. $\Delta_{\rm rms| BH}$= 0.64 dex), whereas, for ETGs it is a bit higher about the $M_{\rm BH}$--$\rm R_{e,sph}$ relation (cf. $\Delta_{\rm rms| BH}$= 0.52 dex about the $M_{\rm BH}$--$M_{\rm *,sph}$ relation).
 
We did not find significantly different relations upon dividing our total sample into S\'ersic versus core-S\'ersic galaxies, or barred versus non-barred galaxies, in the $M_{\rm BH}$--$R_{e,sph}$ diagram. However, when we perform separate regressions for ETGs with a stellar disk (ES-, and S0-types) and ETGs without a stellar disk  (E-type), we find two almost parallel relations which are offset by $\sim\rm1 \, dex$ in the $\log(M_{\rm BH})$-direction (see Figure \ref{MBH-Re_disk}). 
ETGs with a disk follow the relation
\begin{IEEEeqnarray}{rCl}
\label{Re_Mbh_ESS0}
\log(M_{\rm BH}/M_\odot) &=& (2.13\pm 0.22)\log\left(\rm R_{e,sph, maj}/kpc \right) \nonumber \\
&& +\> (8.34\pm 0.09),
\end{IEEEeqnarray}
with $\Delta_{\rm rms|BH} = 0.55$ dex, and  ETGs without a disk define
\begin{IEEEeqnarray}{rCl}
\label{Re_Mbh_E}
\log(M_{\rm BH}/M_\odot) &=& (1.78\pm 0.24)\log\left(\rm R_{e,sph, maj}/kpc \right) \nonumber \\
&& +\> (7.24\pm 0.25),
\end{IEEEeqnarray}
with $\Delta_{\rm rms|BH} = 0.60$ dex.
The $M_{\rm BH}$--$\rm R_{e,sph, eq}$ relations for ETGs with disk and ETGs without a disk, which are consistent with above Equations \ref{Re_Mbh_ESS0} and \ref{Re_Mbh_E}, respectively, are provided in Table \ref{fit parameters2}.  
The two relations defined by ETGs with and without a disk are steeper than the single-regression $M_{\rm BH}$--$\rm R_{e,sph}$ relation for ETGs (Equation \ref{Re_Mbh_ETG}); however, the vertical scatter is comparable. The $M_{\rm BH}$--$\rm R_{e,sph}$ relation for the LTGs (Equation \ref{Re_Mbh_LTG}) is slightly steeper, but still its slope is consistent with the slope of the relations for ETGs with and without a disk at the $1 \sigma$ level; however the intercepts are different. The final substructures in the $M_{\rm BH}$--$\rm R_{e,sph}$ diagram, i.e., the relations followed by ETGs with a disk, ETGs without a disk, and LTGs, are presented together in Figure \ref{MBH-Re_triple}.

In passing, we note that the $M_{\rm BH}$--$\rm R_{e,sph}$ relation that we obtained for ETGs without a disk (elliptical galaxies) ---most of which are core-S\'ersic galaxies---is also consistent with the relation $M_{\rm BH} \propto \rm R_{e}^{1.86 \pm 0.26}$ obtained by combining the $M_{\rm BH}$--(break radius or depleted core radius,  $\rm R_b$) and $\rm R_b$--$\rm R_e$ relations observed for cored galaxies in \citet{Dullo:2014}.

This offset between ETGs with and without a disk in the $M_{\rm BH}$--$\rm R_{e,sph}$ diagram is analogous to the offset found between the parallel relations for ETGs with and without a disk in the $M_{\rm BH}$--$M_{\rm *,sph}$ diagram \citep[][their figure 8]{Sahu:2019:I}. Also, on combining the $M_{\rm BH}$--$M_{\rm *,sph}$ relations defined by ETGs with and without a disk \citep[][their Equations 12 and 13]{Sahu:2019:I},  with our $M_{\rm *,sph}$--$\rm R_{e,sph}$ relation (Equation \ref{Re_Msph}),  we obtain $M_{\rm BH} \propto \rm R_{e,sph, maj}^{2.01 \pm 0.23}$ and $M_{\rm BH} \propto \rm R_{e,sph, maj}^{2.05 \pm 0.23}$ ($M_{\rm BH} \propto \rm R_{e,sph, eq}^{2.05 \pm 0.23}$ and $M_{\rm BH} \propto \rm R_{e,sph, eq}^{2.09 \pm 0.23}$, for $\rm R_{e,sph,eq}$), which are consistent with the observed relations for ETGs with a disk and ETGs without a disk, respectively (Equation \ref{Re_Mbh_ESS0} \& \ref{Re_Mbh_E}, see Table \ref{fit parameters2} for $M_{\rm BH} -\rm R_{e,sph, eq}$ parameters). 

Importantly, as mentioned in \citet{Sahu:2019:I}, this order of magnitude offset has little to do with the black hole masses of these two categories. Qualitatively, this offset can be  understood by the different sizes of the spheroid effective half-light radius corresponding to ETGs with a disk (ES and S0) and ETGs without a disk (E). The ellicular (ES) and lenticular (S0) galaxies, which have intermediate/large-scale stellar disks in addition to their spheroids, have a smaller $\rm R_{e,sph}$ relative to the elliptical galaxies which are comprised (almost) entirely of spheroids.
This difference in $\rm R_{e,sph}$ between the two sub-populations of ETGs creates the offset between the $M_{\rm BH}$--$\rm R_{e,sph}$ relations defined by them, and because of the non-zero slope of the $M_{\rm BH}$--$\rm R_{e,sph}$ relations, we see an offset in the vertical direction. 
 
The relation $M_{\rm *,sph}=(M/L)2 \pi R_e^2 \langle I \rangle_e$  \citep[e.g.\ Equation 8 in][]{Graham:Re:2019}, where (M/L) represents the stellar mass-to-light ratio and $\rm \langle I \rangle_e$ is the averaged intensity within $\rm R_e$, suggests that $\rm \log(M_{\rm *,sph}) \propto \log(\langle I \rangle_e)+2\log(R_e)$, for a constant mass-to-light ratio.
This can help us quantitatively understand the origin of the offset (of $1.12\pm 0.20$ dex) found in the $\log(M_{\rm BH})$--$\log(M_{\rm *,sph})$ diagram \citep[][their section 4.2]{Sahu:2019:I} between ETGs with and without a stellar disk. We find a vertical offset of $1.41 \pm 0.23$~dex between the two sub-samples of ETGs in the $\log(M_{\rm BH})$--$\rm 2\log(R_{e,sph,eq})$ diagram. Whereas, we do not find separate statistically significant $\log(M_{\rm BH})$--$\rm log(\langle I \rangle_e)$ relations for these two populations, implying a single  $\log(M_{\rm BH})$--$\rm log(\langle I \rangle_e)$ for ETGs with and with out a disk.
This suggests that the offset observed in the $\log(M_{\rm BH})$--$\log(M_{\rm *,sph})$ diagram by \citet{Sahu:2019:I}  originates mainly from the offset in the $\log(M_{\rm BH})$--$\rm \log(R_{e,sph})$ diagram.  

Furthermore, in the plot of $M_{\rm BH}$ versus the effective radius of the whole galaxy ($\rm R_{e,gal}$), this offset is expected to disappear, such that all the ETGs will follow a single $M_{\rm BH}$--$\rm R_{e,gal}$ relation, analogous to the combined behaviour of  ETGs with and without a disk in the $M_{\rm BH}$--$M_{\rm *,gal}$ diagram \citep[][see the right-hand panel of their Figure 8]{Sahu:2019:I}, where the two sub-populations of ETGs follow consistent $M_{\rm BH}$--$M_{\rm *,gal}$ relations.

\begin{figure*}
\begin{center}
\includegraphics[clip=true,trim= 06mm 03mm 10mm 10mm,width=   1.0\textwidth]{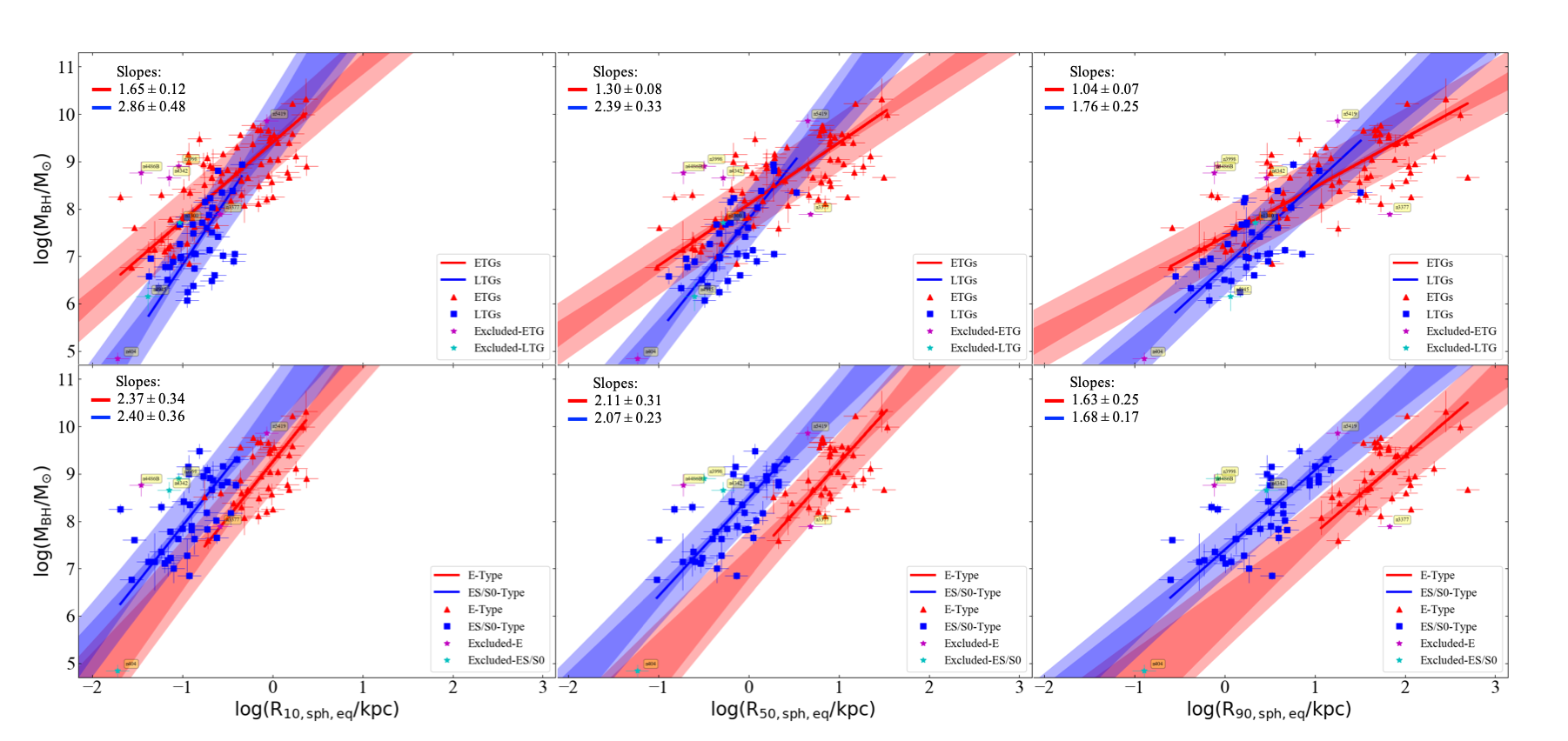}
\caption{Similar to Figure \ref{MBH-Re} and \ref{MBH-Re_disk}, but now also showing the correlations of $M_{\rm BH}$ with the radius containing $z=10\%$ ($\rm R_{10,sph,eq}$) and $z=90 \%$ ($\rm R_{90,sph,eq}$) of the spheroid's light in the left and right panels, in addition to $M_{\rm BH}$--$\rm R_{e,sph,eq}$ relations in the middle panel. Top panels reveal that ETGs and LTGs follow two different relations in all three cases. The bottom panel reveals that the offset between ETGs with and without a disk is obtained in all cases, where the offset varies with z. Additionally, due to the more massive systems having larger S\'ersic indices, for all sub-morphologies the slope of the $M_{\rm BH}$--$\rm R_{z,sph,eq}$ relation gradually decreases with increasing z. Intercepts and the scatter about these relations can be found in Table \ref{fit parameters2}. }
\label{combo_Mbh}
\end{center}
\end{figure*}

Similar to the previous subsection, here also we investigate the correlations of black hole mass with radii containing $\rm z=10 \%$ and $\rm z=90 \%$ of the spheroid's total light, in addition to the $\rm 50 \%$ ($\rm R_{e,sph}$) radius discussed above. Figure \ref{combo_Mbh} presents the correlations we observed between black hole mass and $\rm R_{10,sph,eq}$,  $\rm R_{50,sph,eq}$ (or $\rm R_{e,sph,eq}$), and $\rm R_{90,sph,eq}$, respectively in the left, middle, and right panels. The top panels show that ETGs and LTGs define two different  $M_{\rm BH}$--$\rm R_{z,sph,eq}$ relations irrespective of z, however, the slopes of the relations become shallower with increasing z. The bottom panels reveal that the offset between the $M_{\rm BH}$--$\rm R_{z,sph,eq}$ relations followed by ETGs with a disk and ETGs without a disk are found in all cases; however, as expected, the amount of the offset varies with z and also the slopes of these relations become shallower with increasing z. The parameters for the $M_{\rm BH}$--$\rm R_{z,sph,eq}$ relations obtained for $\rm z=10 \%$ and $\rm z=90 \%$ are also presented in Table \ref{fit parameters2}.

\vspace*{-1.2mm}
\startlongtable
\begin{deluxetable*}{lcrrccrrrr}
\tabletypesize{\footnotesize}
\tablecolumns{10}
\tablecaption{Correlations of $M_{\rm \rm *,sph}$ and $M_{\rm BH}$ with the bulge/spheroid major-axis properties ($\rm n_{sph,maj} \, and \, R_{e,sph,maj}$)\label{fit parameters}}
\tablehead{
\colhead{ \textbf{Category} } & \colhead{ \textbf{Number} } & \colhead{ \bm{$\alpha$} } & \colhead{ \bm{$\beta$ } } & \colhead{ \bm{$\epsilon$} } & \colhead{ \bm{ $\Delta_{\rm rms}$} }  & \colhead{ \bm{$r$ } } & \colhead{ \bm{$\log p $ }} & \colhead{ \bm{$r_s$}} & \colhead{ \bm{$\log p_s $ }}  \\
\colhead{} & \colhead{} & \colhead{} & \colhead{ \textbf{dex}} & \colhead{\textbf{dex}} & \colhead{\textbf{dex}} & \colhead{} & \colhead{ \textbf{dex}} & \colhead{} & \colhead{\textbf{dex}} \\
\colhead{ \textbf{(1)}} & \colhead{\textbf{(2)}} & \colhead{\textbf{(3)}} & \colhead{\textbf{(4)}} & \colhead{\textbf{(5)}} & \colhead{\textbf{(6)}} & \colhead{\textbf{(7)}} & \colhead{\textbf{(8)}} & \colhead{\textbf{(9)}} & \colhead{\textbf{(10)}}
}
\startdata
\multicolumn{10}{c}{ $\log(M_{*,\rm sph}/{\rm M_{\sun}})=\alpha\log(\rm n_{sph,maj}/3)+\beta$} \\
\hline
ETGs & 77 & $ 3.27 \pm 0.25  $ &  $ 10.50\pm 0.06 $ &  0.27 & 0.46  & 0.79 & -17.14 & 0.80 & -17.32 \\
LTGs & 38 & $ 1.31\pm 0.22 $ & $10.41 \pm 0.07  $ & 0.23 & 0.32 & 0.54 & -3.28 & 0.41 &  -2.01 \\
\hline
\multicolumn{10}{c}{ $\log(M_{\rm BH}/{\rm M_{\sun}})=\alpha\log(\rm n_{sph,maj}/3)+\beta$} \\
\hline
All Galaxies &  115 & $ 3.79 \pm 0.23 $ & $8.15 \pm 0.06 $ & 0.60 & 0.69 &  0.77 &  -22.85 & 0.76 & -22.36   \\
ETGs &  77 & $ 3.95\pm 0.34 $ & $8.15 \pm 0.08 $ & 0.54 & 0.65  & 0.71  &  -12.46 & 0.69 & -11.42  \\
LTGs & 38  & $ 2.85\pm 0.31 $ & $ 7.90\pm 0.14 $ &  0.62 & 0.67 & 0.53  & -3.20  & 0.45 & -2.36  \\
Non-Barred & 71  & $ 3.89\pm 0.31 $ & $ 8.15\pm 0.10 $ &  0.61 & 0.70 & 0.72  & -11.80 & 0.64 & -8.61  \\
Barred &  44 & $ 3.08\pm 0.36 $ & $8.00 \pm 0.11 $ & 0.55 & 0.61  & 0.54  &  -3.76 & 0.47 & -2.95  \\
\hline
\multicolumn{10}{c}{ $\log(M_{*,\rm sph}/{\rm M_{\sun}})=\alpha\log(\rm R_{e,sph, maj})+\beta$} \\
\hline
All Galaxies &  115 & $ 1.08 \pm 0.04 $ & $10.32 \pm 0.03 $ & 0.13 & 0.27 &  0.92 &  -47.92 & 0.90 & -42.83   \\
ETGs & 77  & $1.03 \pm 0.05 $ & $10.39 \pm 0.04 $ & 0.09  & 0.25  & 0.93  &  -34.73 & 0.94 & -36.98 \\
LTGs & 38  & $1.01 \pm 0.15 $ & $10.24 \pm 0.05 $ & 0.16  & 0.27  & 0.66  &    -5.26 & 0.58 & -3.84  \\
\hline
\multicolumn{10}{c}{ $\log(M_{\rm BH}/{\rm M_{\sun}})=\alpha\log(\rm R_{e,sph,maj})+\beta$} \\
\hline
ETGs with a disk & 39  & $ 2.13\pm 0.22 $ & $8.34 \pm 0.09 $ &  0.47 & 0.55  &  0.74  &   -7.21 & 0.76 & -7.81 \\
ETGs without a disk & 38 & $1.78 \pm 0.24 $ & $7.24 \pm 0.25 $ &  0.55 & 0.60   &  0.55  &  -3.45 & 0.52 & -3.13 \\
ETGs (All) & 77 & $1.26 \pm 0.08 $ & $8.00 \pm 0.07 $ &  0.54 & 0.58   &  0.76  &    -15.10 & 0.74 & -13.95 \\
LTGs & 38  & $ 2.33\pm 0.31 $ & $7.54 \pm 0.10 $ &  0.54 & 0.62  &  0.63  &   -4.66 & 0.62 & -4.43 \\
All Galaxies & 115 & $1.59 \pm 0.09 $ & $7.73 \pm 0.07 $ & 0.63 & 0.67 & 0.78 & -23.56  & 0.78 & -24.26 \\
\enddata
\tablecomments{
Columns:
(1) Subclass of galaxies.
(2) Number of galaxies in a subclass.
(3) Slope of the line obtained from the \textsc{BCES(Bisector)} regression.
(4) Intercept of the line obtained from the \textsc{BCES(Bisector)} regression.
(5) Intrinsic scatter in the vertical ($\log M_{\rm *,sph}$ or $\log M_{\rm BH}$)-direction \citep[see Equation 1 from][]{Graham:Driver:2007}.
(6) Total root mean square (rms) scatter in the vertical direction.
(7) Pearson correlation coefficient.
(8) Pearson correlation probability value.
(9) Spearman rank-order correlation coefficient.
(10) Spearman rank-order correlation probability value.
}
\end{deluxetable*}

\startlongtable
\begin{deluxetable*}{lcrrccrrrr}
\tabletypesize{\footnotesize}
\tablecolumns{10}
\tablecaption{Correlations of $M_{\rm *,sph}$ and $M_{\rm BH}$ with the bulge/spheroid equivalent-axis properties ($\rm n_{eq,sph},  C(1/3), R_{e,sph,eq}, R_{10,sph,eq}, and \, R_{90,sph,eq}$)\label{fit parameters2}}
\tablehead{
\colhead{ \textbf{Category} } & \colhead{ \textbf{Number} } & \colhead{ \bm{$\alpha$} } & \colhead{ \bm{$\beta$ } } & \colhead{ \bm{$\epsilon$} } & \colhead{ \bm{ $\Delta_{\rm rms}$} }  & \colhead{ \bm{$r$ } } & \colhead{ \bm{$\log p $ }} & \colhead{ \bm{$r_s$}} & \colhead{ \bm{$\log p_s $ }}  \\
\colhead{} & \colhead{} & \colhead{} & \colhead{ \textbf{dex}} & \colhead{\textbf{dex}} & \colhead{\textbf{dex}} & \colhead{} & \colhead{ \textbf{dex}} & \colhead{} & \colhead{\textbf{dex}} \\
\colhead{ \textbf{(1)}} & \colhead{\textbf{(2)}} & \colhead{\textbf{(3)}} & \colhead{\textbf{(4)}} & \colhead{\textbf{(5)}} & \colhead{\textbf{(6)}} & \colhead{\textbf{(7)}} & \colhead{\textbf{(8)}} & \colhead{\textbf{(9)}} & \colhead{\textbf{(10)}}
}
\startdata
\multicolumn{10}{c}{ $\log(M_{*,\rm sph}/{\rm M_{\sun}})=\alpha\log(\rm n_{sph,eq}/3)+\beta$} \\
\hline
ETGs & 77 & $ 3.34 \pm 0.24  $ &  $ 10.52\pm 0.06 $ &  0.32 & 0.49  & 0.77 & -15.63 & 0.74 & -13.86 \\
LTGs & 38 & $1.37\pm 0.20 $ & $10.46 \pm 0.07  $ & 0.20 & 0.28 & 0.63 & -4.61 & 0.55 &  -3.39 \\
\hline
\multicolumn{10}{c}{ $\log(M_{\rm BH}/{\rm M_{\sun}})=\alpha\log(\rm n_{sph,eq}/3)+\beta$} \\
\hline
All Galaxies &  115 & $ 3.72 \pm 0.23 $ & $8.20 \pm 0.07 $ & 0.65 & 0.73 &  0.74 &  -20.14 & 0.74 & -20.22  \\
ETGs &  77 & $ 3.94\pm 0.37 $ & $8.18 \pm 0.09 $ & 0.63 & 0.73  & 0.64  &  -9.37 & 0.60 & -8.13 \\
LTGs & 38  & $ 2.86\pm 0.33 $ & $ 7.99\pm 0.16 $ &  0.64 & 0.68 & 0.49  & -2.78  & 0.50 & -2.86  \\
Non-Barred &  71 & $ 4.20\pm 0.38 $ & $8.11 \pm 0.10 $ & 0.69 & 0.79  & 0.64  &  -8.89 & 0.54 & -5.88 \\
Barred & 44  & $ 2.91\pm 0.30 $ & $ 8.08\pm 0.13 $ &  0.56 & 0.61 & 0.52  & -3.54 & 0.45 & -2.66  \\
\hline
\multicolumn{10}{c}{ $\log(M_{\rm BH}/{\rm M_{\sun}})=\alpha \,  \rm C(1/3)/0.4+\beta$} \\
\hline
All Galaxies &  115 & $ 8.81 \pm 0.53 $ & $8.10 \pm 0.07 $ & 0.65 & 0.73 &  0.74 &  -20.16 & 0.74 & -20.17   \\
ETGs &  77 & $ 8.94 \pm 0.86 $ & $8.10 \pm 0.09 $ & 0.63 & 0.72 &  0.64 &  -9.45 & 0.60 & -8.13   \\
LTGs &  38 & $ 6.88 \pm 0.97 $ & $7.91 \pm 0.16 $ & 0.64 & 0.68 &  0.47 &  -2.57 & 0.50 & -2.83   \\
Non-Barred &  71 & $ 9.75 \pm 0.93 $ & $8.00 \pm 0.11 $ & 0.70 & 0.79 &  0.64 &  -8.77 & 0.54 & -5.88   \\
Barred &  44 & $ 7.03 \pm 0.88 $ & $8.02 \pm 0.13 $ & 0.56 & 0.61 &  0.51 &  -3.36 & 0.45 & -2.63   \\
\hline
\multicolumn{10}{c}{ $\log(M_{*,\rm sph}/{\rm M_{\sun}})=\alpha\log(\rm R_{e,sph,eq})+\beta$} \\
\hline
All Galaxies &  115 & $ 1.10 \pm 0.04 $ & $10.42 \pm 0.03 $ & 0.08 & 0.26 &  0.93 &  -50.60 & 0.92 & -46.02   \\
ETGs & 77  & $1.06 \pm 0.05 $ & $10.46 \pm 0.04 $ & 0.08  & 0.26  & 0.93  &  -33.85  & 0.94 & -35.31 \\
LTGs & 38  & $1.03 \pm 0.12 $ & $10.34 \pm 0.05 $ & 0.00  & 0.22  & 0.78  &  -7.98 & 0.67 & -5.39 \\
\hline
\multicolumn{10}{c}{ $\log(M_{\rm BH}/{\rm M_{\sun}})=\alpha\log(\rm R_{e,sph,eq})+\beta$} \\
\hline
ETGs with a disk & 39 & $2.07 \pm 0.23 $ & $8.49 \pm 0.09 $ &  0.52 & 0.59   &  0.70  &  -6.19 & 0.71 & -6.33 \\
ETGs without a disk & 38  & $ 2.11\pm 0.31 $ & $7.11 \pm 0.27 $ &  0.55 & 0.61  &  0.53  &  -3.27 & 0.46 & -2.43 \\
ETGs & 77 & $1.30 \pm 0.08 $ & $8.10 \pm 0.07 $ &  0.56 & 0.60   &  0.75  &  -14.41  & 0.72 & -12.76 \\
LTGs & 38  & $ 2.39\pm 0.33 $ & $7.79 \pm 0.13 $ &  0.52 & 0.60  &  0.66  &  -5.13 & 0.66 & -5.21 \\
All Galaxies & 115 & $1.62 \pm 0.09 $ & $7.86 \pm 0.06 $ & 0.62 & 0.67 & 0.78 & -23.81  & 0.78 & -24.29 \\
\hline
\multicolumn{10}{c}{ $\log(M_{*,\rm sph}/{\rm M_{\sun}})=\alpha\log(\rm R_{10,sph,eq})+\beta$} \\
\hline
All Galaxies &  115 & $ 1.47 \pm 0.06 $ & $11.51 \pm 0.04 $ & 0.17 & 0.33 &  0.88 &  -38.55 & 0.86 & -34.27   \\
\hline
\multicolumn{10}{c}{ $\log(M_{\rm BH}/{\rm M_{\sun}})=\alpha\log(\rm R_{10,sph,eq})+\beta$} \\
\hline
ETGs with a disk & 39 & $2.39 \pm 0.36 $ & $10.30 \pm 0.32 $ &  0.60 & 0.68   &  0.61  &  -4.38 & 0.63 & -4.78 \\
ETGs without a disk & 38  & $ 2.37\pm 0.34 $ & $9.25 \pm 0.10 $ &  0.53 & 0.61  &  0.54  &  -3.37 & 0.47 & -2.54 \\
ETGs & 77 & $1.65 \pm 0.12 $ & $9.40 \pm 0.09 $ &  0.56 & 0.62  &  0.73  &  -13.50  & 0.71 & -12.24 \\
LTGs & 38  & $ 2.86\pm 0.47 $ & $9.67 \pm 0.44 $ &  0.60 & 0.69  &  0.54  &  -3.30 & 0.50 & -2.83 \\
\hline
\multicolumn{10}{c}{ $\log(M_{*,\rm sph}/{\rm M_{\sun}})=\alpha\log(\rm R_{90,sph,eq})+\beta$} \\
\hline
All Galaxies &  115 & $ 0.85 \pm 0.03 $ & $9.90 \pm 0.03 $ & 0.12 & 0.26 &  0.93 &  -50.53 & 0.92 & -47.05   \\
\hline
\multicolumn{10}{c}{ $\log(M_{\rm BH}/{\rm M_{\sun}})=\alpha\log(\rm R_{90,sph,eq})+\beta$} \\
\hline
ETGs with a disk & 39 & $1.68 \pm 0.17 $ & $7.40 \pm 0.13 $ &  0.51 & 0.56   &  0.73  &  -6.89 & 0.71 & -6.35 \\
ETGs without a disk & 38  & $ 1.63 \pm 0.25 $ & $6.11 \pm 0.42 $ &  0.61 & 0.63  &  0.47  &  -2.52 & 0.41 & -1.95 \\
ETGs & 77 & $1.04 \pm 0.07 $ & $7.41 \pm 0.09 $ &  0.58 & 0.60   &  0.74  &  -14.09  & 0.71 & -12.17 \\
LTGs & 38  & $ 1.76\pm 0.25 $ & $6.78 \pm 0.10 $ &  0.54 & 0.58  &  0.67  &  -5.30 & 0.68 & -5.50 \\
\enddata
\tablecomments{Column names are same as Table \ref{fit parameters}.
}
\end{deluxetable*}

\section{\textbf{Summary}}
\label{summary}
We have used the largest sample of galaxies to date with directly-measured black hole masses, and carefully measured bulge parameters obtained from multi-component decomposition of their galaxy light in our previous studies \citep{Savorgnan:Graham:2016:I, Davis:2018:a, Sahu:2019:I}. Using this extensive data-set, we have investigated the correlations between black hole mass ($M_{\rm BH}$) and the bulge S\'ersic index ($\rm n_{sph}$), bulge central light concentration index (C), and the bulge effective half-light radius ($\rm R_{e,sph}$). 

For our sample, we also investigated the correlations between bulge mass ($M_{\rm *, sph}$) and both the bulge S\'ersic index and bulge half-light radius. We then combined these with the latest $M_{\rm BH}$--$M_{*, sph}$ relations to  predict and check upon the observed correlations of $M_{\rm BH}$ with $\rm n_{sph}$ and $\rm R_{e,sph}$. 

In all of the relations we investigated, we explored the possibility of substructure due to various subcategories of galaxy morphology, i.e., S\'ersic versus core-S\'ersic galaxies, galaxies with a stellar disk versus galaxies without a stellar disk, barred versus non-barred galaxies, and ETGs versus LTGs.  

Parameters for all the correlations presented  in this paper are separately listed in Table \ref{fit parameters} and Table \ref{fit parameters2}. The slope of the correlations that we obtained for $M_{\rm BH}$ or $M_{\rm *, sph}$ with the major-axis bulge parameters ($\rm n_{sph,maj}$ and $\rm R_{e,sph, maj}$) are consistent with the slope from the corresponding correlations of $M_{\rm BH}$ or $M_{*, sph}$ with the equivalent-axis bulge parameters ($\rm n_{sph,eq}$ and $\rm R_{e,sph, eq}$). 

Our prime results can be summarized as follows,

\begin{itemize}
\item{ETGs and LTGs follow two different $M_{\rm *,sph}$--$\rm n_{sph}$ relations  (see Figure \ref{Msph-nmaj}),  with slopes equal to $3.27\pm0.25$ and $1.31\pm0.22$, and total rms scatter equal to $\Delta_{\rm rms|sph}$=0.46~dex and 0.32~dex, respectively (Equations \ref{n_Msph_ETGs} and \ref{n_Msph_LTGs}), in the $M_{\rm *,sph}$--$\rm n_{sph,maj}$ diagram. As  the S\'ersic index is a measure of the central concentration of a bulge's light, these different slopes for the $M_{\rm *,sph}$--$\rm n_{sph}$  relation suggest  distinct mechanisms for the evolution of spheroid mass and central light (or stellar mass) concentration in ETGs and LTGs.}

\item{In the $M_{\rm BH}$--$\rm n_{sph}$ diagram, ETGs and LTGs seem to follow two different relations with $M_{\rm BH} \propto \rm n_{sph, maj}^{3.95\pm 0.34}$  and $M_{\rm BH} \propto \rm n_{sph, maj}^{2.85\pm 0.31}$  with $\Delta_{\rm rms|BH}$ = 0.65 dex and 0.67 dex, respectively (Figure \ref{Mbh-nmaj2}, Equations \ref{n_Mbh_ETG} and \ref{n_Mbh_LTG}).}

\item{In the diagram showing the black hole mass versus the spheroid central concentration index, C(1/3),  we again find two (slightly) different relations due to ETGs and LTGs (Figure \ref{Mbh-CI2}, Table \ref{fit parameters2}), analogous to the $M_{\rm BH}$--$\rm n_{sph}$ diagram. The slopes for the $M_{\rm BH}$--$\rm C(1/3)$ relations are $8.94\pm0.86$ and $6.88\pm0.97$ with $\Delta_{\rm rms|BH}$= 0.72~dex and 0.68~dex, respectively, for ETGs and LTGs.
}

\item{We find a tight near-linear relation between  $M_{\rm *,sph}$ and $\rm R_{e,sph}$  for our range  of data (Figures \ref{Msph-Re} and \ref{Msph-Re_single}). Both ETGs and LTGs define the log-linear relation $M_{\rm *,sph} \propto \rm R_{e,sph,maj}^{1.08\pm 0.04}$ (Equation \ref{Re_Msph}) with $\Delta_{\rm rms|sph}$=0.27 dex. An extended view of the $M_{\rm *,gal}$--$\rm R_{e,gal}$  relation for ETGs  is curved \citep{Graham:Re:2019}, and our $M_{\rm *,sph}$--$\rm R_{e,sph}$ relation, somewhat dominated by massive spheroids, agrees with the quasi-linear part of the curve at high-masses where E-type galaxies dominate.}

\item{ETGs and LTGs define two different relations between black hole mass and bulge $\rm R_e$ (Figure \ref{MBH-Re}), such that $M_{\rm BH}$--$\rm R_{e,sph, maj}^{(1.26\pm 0.08)}$ and $M_{\rm BH}$--$\rm R_{e,sph, maj}^{(2.33\pm 0.31)}$ for ETGs and LTGs, with $\Delta_{\rm rms|BH}$=0.58 dex and  0.62 dex, respectively (Equation \ref{Re_Mbh_ETG} and \ref{Re_Mbh_LTG}). This is analogous to the substructure in the $M_{\rm BH}$--$M_{\rm *,sph}$ diagram due to ETGs and LTGs \citep{Sahu:2019:I}.}

\item{ In the $M_{\rm BH}$--$\rm R_{e,sph}$ diagram, ETGs with a disk (ES, S0) and ETGs without a disk (E) follow two different, almost parallel, relations with slopes $\sim 2\pm 0.2$ (Figure \ref{MBH-Re_disk}), which are steeper than the above single-regression $M_{\rm BH}$--$\rm R_{e,sph}$ relation for all ETGs (see Tables \ref{fit parameters} and \ref{fit parameters2} for parameters) and offset by a factor of $\sim$10 in the vertical $M_{\rm BH}$-direction. This is again analogous to the offset observed between the $M_{\rm BH}$--$M_{\rm *,sph}$ relations followed by ETGs with and without a disk \citep{Sahu:2019:I}. 
Given $M_{\rm *,sph}$ depends  on $\rm R_{e,sph}$ via $M_{\rm *,sph}=(M/L)2\pi R_e^2 \langle I \rangle_e$, we find that the offset in the $M_{\rm BH}$--$M_{\rm *,sph}$ diagram originates from the offset between ETGs with and without a disk in the $M_{\rm BH}$--$\rm R_{e,sph}$ diagram. The  reason behind the offset is the smaller spheroid half-light radius of ETGs with a disk relative to that of elliptical (purely spheroidal) galaxies.}

\item{ In the $M_{\rm *,sph}$--$\rm R_{z,sph}$ and $M_{\rm BH}$--$\rm R_{z,sph}$  diagrams for z=$10\%$ and $90 \%$ (see Figures \ref{combo_Msph} and \ref{combo_Mbh}),  we recover the same substructures as the $M_{\rm BH}$--$\rm R_{e,sph}$ and $M_{\rm *,sph}$--$\rm R_{e,sph}$ relations mentioned above, with the slopes of correlations gradually decreasing with increasing z (see Table \ref{fit parameters2} for parameters).  }

\end{itemize}

The $M_{\rm BH}$--$\rm n_{sph}$ and $M_{\rm BH}$--$\rm R_{e,sph}$ relations may be useful for predicting the black hole masses of galaxies using their bulge S\'ersic index or bulge half-light radius parameters. These parameters can be obtained by performing a multi-component decomposition of a galaxy light profile obtained even  from a photometrically uncalibrated image. 
One should be careful while using the $M_{\rm BH}$--$\rm R_{e,sph}$ relation, because  ETGs  with a disk (ES,S0), ETGs without a disk (E), and LTGs (spirals) are found to follow different trends (Figures \ref{MBH-Re} and \ref{MBH-Re_disk}). However, when extended ETG or LTG classification is not known, the single regression $M_{\rm BH}$--$\rm n_{sph}$  or  $M_{\rm BH}$--$\rm R_{e,sph}$ relations (provided in Tables \ref{fit parameters} and \ref{fit parameters2}) can still be used to predict $M_{\rm BH}$, albeit with a higher uncertainty.

Our BH scaling relations, based on local galaxies, form a benchmark for  studies investigating the evolution of BH correlations with galaxy properties across cosmic time \citep{Lapi:Raimundo:2014, Park:Woo:2015, Sexton:2019, Suh:Civano:2020}. In addition to enabling one to determine the black hole mass function \citep[e.g.][]{McLure:Dunlop:2004, Shankar:Salucci:2004, Graham:MGC:2007, Vika:Driver:2009, Davis:Berrier:2014, MutluPakdil:Seigar:2016}, these BH scaling relations with bulge S\'ersic parameters can also be employed to infer the lifetime of binary black holes \citep{Biava2019, Li:Ballantyne:2020} and further constrain the BH merger rate.
The creation of merger-built spheroids with (initially) higher central stellar densities --- which are associated with higher S\'ersic indices --- should, through dynamical friction \citep[e.g.,][]{Chandrasekhar:1943, ArcaSedda:2014}, experience a quicker inspiral and hardening phase for their binary black holes.  The imprint of such processes are the phase-space loss-cones \citep{Begelman:1980} observed as partially-depleted cores in massive spheroids \citep{King:Minkowski:1966, King:Minkowski:1972,  Lauer:1985, Ferrarese:vandenBosch:1994, Trujillo:2004:A20, Dullo:2014}.  The eventual coalescence of the black holes results in the emission of gravitational waves \citep{Poincare:1906, Einstein:1916,  Einstein:1918, Abbott:GW:2016}. 
Our BH scaling relations will play a key role in constraining the detection of low-frequency gravitational waves generated from BH mergers at high redshifts, which fall in the detection domain of pulsar timing arrays  \citep{Shannon:2015, Lentati:2015, Sesana2016, Arzoumanian:2018} and LISA \citep{Amaro-Seoane:Audley:2017,Barack2019}.

The different scaling relations for ETGs and LTGs also hold valuable information for simulations, analytical/semi-analytical, and  theoretical models of galaxy formation and evolution \citep[e.g.][]{Volonteri:2013, Heckman:2014, Conselice2014}, as they reveal the  trends of BH---host bulge/galaxy properties depending on galaxy morphology. These relations can be used for primary size and structure tests in simulations aiming to generate realistic galaxies with supermassive black holes at their center \citep[e.g.][]{SchayeEagle2015, HopkinsFIRE2018, MutluPakdil:Seigar:2018, Dave:Simba:2019, LiYuan2020}. We plan to test  our new constraints through a comparison with simulations in our future work. 
Using our extensive dataset, we will present the correlation of black hole mass with the internal stellar density of galactic spheroids (N. Sahu et al. 2021, in preparation). We will also explore the (first morphology aware) fundamental plane in our future work.

\acknowledgements
We thank the anonymous referee whose comments helped us improve the clarity of this paper. This research was conducted with the Australian Research Council Centre of Excellence for Gravitational Wave Discovery (OzGrav), through project number CE170100004. This project was supported under the Australian Research Council's funding scheme DP17012923.
This work has made use of the NASA/IPAC Infrared Science Archive, the NASA/IPAC Extragalactic Database (NED), and the \textsc{HyperLeda} Database \url{http://leda.univ-lyon1.fr/}.

\bibliography{bibliography}

\appendix
\section{Data Set}
\label{appendix_data}
In Table \ref{Total Sample}, first 83 galaxies are ETGs, and the remaining are LTGs, where the galaxies with a depleted core are marked with superscript \enquote{a} on their names in the first column.  
The spheroid S\'ersic model parameters ($\rm n, Re, \mu$), morphology, and spheroid stellar masses are taken from our previous studies \citet{Savorgnan:Graham:2016:I}, \citet{Sahu:2019:I}, and \citet{Davis:2018:a}. For NGC~1271 and NGC~1277 these parameters are borrowed from \citet{Graham:Ciambur:Savorgnan:2016} and \citet{Graham:Durr:Savorgnan:2016}, respectively. Spheroid parameters for the Milky Way are taken from \citet{Graham:Driver:2007} who used the uncalibrated bulge surface brightness profile of Milky Way from \citet{Kent:Dame:1991}. The spheroid mass of Milky Way is taken from \citet{LicquiaNewman2015}.

Correlations of $M_{\rm *,sph}$ and $M_{\rm BH}$ with the equivalent-axis bulge parameters obtained using symmetric \textsc{MPFITEXY} regression are presented in Table \ref{fit parameters3}. These relations are consistent with the corresponding relations obtained using the (bisector) \textsc{BCES} regression presented in Table \ref{fit parameters2}.

\restartappendixnumbering
\startlongtable
\movetableright=-2in
\movetabledown=0.5in
\begingroup
\setlength{\tabcolsep}{1.5pt} 
\begin{longrotatetable}
\begin{deluxetable*}{lllcccccclrcclrrclr}
\tablecolumns{19}
\tablecaption{Galaxy Sample}
\tabletypesize{\scriptsize} 
\tablehead{
\colhead{\textbf{No.}} & \colhead{ \textbf{Galaxy} } & \colhead{ \textbf{Band}} & \colhead{ \bm{$\Upsilon_\lambda \,\&\, \mathfrak{M}_{\odot, \lambda}$} } & \textbf{Type} &  \colhead{ \bm{$n_{maj}$}} &  \colhead{ \bm{$R_{e,maj}$}} &  \colhead{ \bm{$\mu_{e,maj}$}} &  \colhead{ \bm{$ \log (\frac{ I_{e,maj}}{M_\odot/pc^2})$}} &  \colhead{ \bm{$n_{eq}$}} & \colhead{ \bm{$R_{e,eq}$}} & \colhead{ \bm{$\mu_{e, eq}$}} & \colhead{ \bm{$\log (\frac{ I_{e,eq}}{M_\odot/pc^2})$}} &  \colhead{ \textbf{C}} &  \colhead{\textbf{scale}}  &  \colhead{\bm{$ \log(\frac{M_{\rm *,sph}}{M_{\odot} })$}} &    \colhead{\textbf{Distance}} & \colhead{\bm{$ \log( \frac{M_{\rm BH} }{M_{\odot} })$}} & \colhead{\textbf{Ref.}} \\
\colhead{} & \colhead{} &\colhead{} & \colhead{$\rm (\frac{M_{\odot}}{L_{\odot}})\, \& \,mag$} & \colhead{}  & \colhead{} & \colhead{arcsec} & \colhead{$\rm \frac{mag}{arcsec^2}$} & \colhead{dex}    & \colhead{} & \colhead{arcsec} & \colhead{ $\rm \frac{mag}{arcsec^2}$}  & \colhead{dex} & \colhead{} & \colhead{$\rm \frac{kpc}{arcsec}$} & \colhead{ dex } &\colhead{Mpc} & \colhead{ dex }  & \colhead{} \\
\colhead{} &\colhead{(1)}  & \colhead{(2)}  & \colhead{(3)} & \colhead{(4)} & \colhead{(5)} & \colhead{(6)} & \colhead{(7)} & \colhead{(8)} & \colhead{(9)} & \colhead{(10)} &\colhead{(11)} & \colhead{(12)}  & \colhead{(13)} & \colhead{(14)} & \colhead{(15)} & \colhead{(16)} & \colhead{(17)} & \colhead{(18)}
} 
\startdata
1 & IC 1459\tablenotemark{a}  & $\rm 3.6\,\mu m$  & (0.6, 3.26) & E & 6.60 & 63.10 & 18.49 & 2.32  & 7.00 & 57.30 & 18.59   & 2.28  & 0.60 & 0.1366 &  11.55 $\pm$  0.12  & 28.4 & 9.38 $\pm$ 0.20  & SG16 \\
2 & NGC 0821 & $\rm 3.6\,\mu m$  & (0.6, 3.26) &  E & 5.30 & 36.50 & 18.40 & 2.36 & 6.10 & 18.90 & 17.83   &  2.58 & 0.62 & 0.1127 &  10.69 $\pm$  0.33  & 23.4 & 7.59 $\pm$ 0.17 & SG16  \\
3 & NGC 1023 & $\rm 3.6\,\mu m$ & (0.6, 3.26)   & SB0  & 2.10 & 9.20 & 14.96 & 3.73  &  2.00 & 7.40 & 14.79   & 3.80 & 0.32 & 0.0536 &  10.21 $\pm$  0.12  & 11.1 & 7.62 $\pm$ 0.05 & SG16   \\
4 & NGC 1316 & $\rm 3.6\,\mu m$ & (0.6, 3.26)  & $\rm SAB0^*$  & 2.00 & 21.50 & 15.55 & 3.50  & 1.80 & 15.90 & 15.43   & 3.54  & 0.30 & 0.0897 &  11.05 $\pm$  0.26  & 18.6 & 8.18 $\pm$ 0.26 & SG16 \\
5 & NGC 1332  & $\rm 3.6\,\mu m$ & (0.6, 3.26)  & ES & 5.10 & 34.70 & 17.44 & 2.74  & 3.70 & 18.00 & 16.47   & 3.13  & 0.35 & 0.1074 &  11.05 $\pm$  0.33  &  22.3 & 9.16$\pm$0.07 & SG16 \\
6 & NGC 1399\tablenotemark{a} & $\rm 3.6\,\mu m$ & (0.6, 3.26)   & E  & 10.00 & 405.10 & 21.80 & 1.00 & 10.00 & 338.10 & 21.53   &  1.10 & 0.63 & 0.0935 &  11.66 $\pm$  0.12  & 19.4  & 8.67$\pm$0.06 & SG16 \\
7 & NGC 2549  & $\rm 3.6\,\mu m$ & (0.6, 3.26)  & SB0  & 2.30 & 6.10 & 15.57 & 3.49 & 1.50 & 3.10 & 14.54   & 3.90  & 0.24 & 0.0594 &  9.59 $\pm$  0.12  & 12.3 & 7.15$\pm$0.6 & SG16 \\
8 & NGC 2778  & $\rm 3.6\,\mu m$ & (0.6, 3.26)   & SAB0 & 1.30 & 2.30 & 15.61 & 3.47  & 1.20 & 2.20 & 15.46   &  3.53 & 0.24 & 0.1074 &  9.41 $\pm$  0.26  & 22.3  & 7.18$\pm$0.34 & SG16 \\
9 & NGC 3091\tablenotemark{a}  & $\rm 3.6\,\mu m$ & (0.6, 3.26)  & E & 7.60 & 100.50 & 20.43 & 1.55 & 6.60 & 51.20 & 19.47   & 1.94  & 0.49 & 0.2447 &  11.61 $\pm$  0.12  &  51.2 & 9.56$\pm$0.04 & SG16 \\
10 & NGC 3115  & $\rm 3.6\,\mu m$ & (0.6, 3.26)  & S0 & 4.40 & 43.60 & 16.67  & 3.04  & 5.10 & 34.40 & 16.85   &  2.97 & 0.58 & 0.0455 &  10.77 $\pm$  0.12  & 9.4 & 8.94$\pm$0.25 & SG16 \\
11 & NGC 3245   & $\rm 3.6\,\mu m$ & (0.6, 3.26)  & SAB0 & 2.90 & 4.40 & 14.96 & 3.73  & 1.70 & 2.40 & 14.00   &  4.12 & 0.24 & 0.0979 &  10.06 $\pm$  0.12  & 20.3 & 8.30 $\pm$0.12 & SG16 \\
12 & NGC 3377   & $\rm 3.6\,\mu m$ & (0.6, 3.26)  & E & 7.70 & 61.80 & 19.16 & 2.05  & 9.20 & 91.70 & 20.33   &  1.58 & 0.73 & 0.0527 &  10.48 $\pm$  0.26  & 10.9 & 7.89 $\pm$0.04 & SG16 \\
13 & NGC 3379\tablenotemark{a}  & $\rm 3.6\,\mu m$ & (0.6, 3.26)  & E  & 5.20 & 57.20 & 17.93   & 2.54  & 5.30 & 50.90 & 17.84   & 2.58  & 0.53 & 0.0498 &  10.8 $\pm$  0.26  & 10.3 & 8.60 $\pm$0.12 & SG16 \\
14 & NGC 3384   & $\rm 3.6\,\mu m$ & (0.6, 3.26)  & SAB0 & 1.60 & 5.50 & 14.21 &  4.03 & 1.80 & 5.60 & 14.56   & 3.89  & 0.32 & 0.0546 &  10.06 $\pm$  0.12  & 11.3  & 7.23 $\pm$0.05 & SG16 \\
15 & NGC 3414  & $\rm 3.6\,\mu m$ & (0.6, 3.26)  & E & 4.80 & 28.00 & 18.10 & 2.48  & 4.50 & 25.50 & 18.08   & 2.48  & 0.46 & 0.118 &  10.83 $\pm$  0.12  & 24.5  & 8.38  $\pm$0.06 & SG16 \\
16 & NGC 3489  & $\rm 3.6\,\mu m$ & (0.6, 3.26)  & SB0  & 1.50 & 2.20 & 13.47   & 4.33  & 1.30 & 1.70 & 13.25   &  4.41 & 0.25 & 0.0565 &  9.54 $\pm$  0.26  & 11.7 & 6.76 $\pm$0.07 & SG16 \\
17 & NGC 3585  & $\rm 3.6\,\mu m$ & (0.6, 3.26)  & E & 5.20 & 105.00 & 19.13   & 2.06 & 6.30 & 86.30 & 19.24   &  2.02 & 0.65 & 0.094 &  11.3 $\pm$  0.26  & 19.5 & 8.49 $\pm$0.13 & SG16 \\
18 & NGC 3607  & $\rm 3.6\,\mu m$ & (0.6, 3.26)  & E   & 5.50 & 69.30 & 19.00   & 2.12 & 5.60 & 65.50 & 19.01   & 2.11  & 0.54 & 0.107 &  11.23 $\pm$  0.26  & 22.2 & 8.11 $\pm$0.18 & SG16 \\
19 & NGC 3608\tablenotemark{a}  & $\rm 3.6\,\mu m$ & (0.6, 3.26)  & E  & 5.20 & 47.50 & 18.93   & 2.14  & 5.70 & 43.40 & 19.00  & 2.12 & 0.58 & 0.1074 &  10.89 $\pm$  0.26  & 22.3 & 8.30 $\pm$0.18 & SG16 \\
20 & NGC 3842\tablenotemark{a}  & $\rm 3.6\,\mu m$ & (0.6, 3.26)  & E  & 8.10 & 100.70 & 21.43   & 1.16  & 8.20 & 73.60 & 21.07   & 1.31 & 0.61 & 0.4643 &  11.92 $\pm$  0.12  & 98.4  & 9.99  $\pm$ 0.13 & SG16 \\
21 & NGC 3998  & $\rm 3.6\,\mu m$ & (0.6, 3.26)  & SAB0 & 1.20 & 5.80 & 15.15   & 3.65  & 1.30 & 4.80 & 14.63   & 3.86  & 0.26 & 0.0662 &  10.02 $\pm$  0.33  & 13.7  & 8.91 $\pm$0.11 & SG16 \\
22 & NGC 4261\tablenotemark{a}  & $\rm 3.6\,\mu m$ & (0.6, 3.26)  & E  & 4.70 & 52.60 & 18.58   & 2.29  & 4.30 & 47.30 & 18.53   & 2.31  & 0.45 & 0.1481 &  11.38 $\pm$  0.26  & 30.8 & 8.70$\pm$ 0.09 & SG16 \\
23 & NGC 4291\tablenotemark{a}  & $\rm 3.6\,\mu m$ & (0.6, 3.26)  & E  & 4.20 & 15.00 & 17.14   &  2.86 & 5.90 & 15.40 & 17.51   &  2.71 & 0.70 & 0.1228 &  10.71 $\pm$  0.26  & 25.5 & 8.52$\pm$ 0.36 & SG16 \\
24 & NGC 4374\tablenotemark{a}   & $\rm 3.6\,\mu m$ & (0.6, 3.26)  & E   & 7.80 & 101.60 & 19.01  & 2.11  & 7.90 & 129.8 & 19.57   & 1.89   & 0.60 & 0.0864 &  11.49 $\pm$  0.26  & 17.9 & 8.95$\pm$ 0.05 & SG16 \\
25 & NGC 4459  & $\rm 3.6\,\mu m$ & (0.6, 3.26)  & S0  & 3.10 & 18.40   &16.69  & 3.04  & 2.60 & 13.00 & 16.23   &  3.22 & 0.34 & 0.0758 &  10.48 $\pm$  0.26  & 15.7 & 7.83$\pm$ 0.09 & SG16 \\
26 & NGC 4472\tablenotemark{a}  & $\rm 3.6\,\mu m$ & (0.6, 3.26)  & E   & 6.60 & 190.20 & 19.33   & 1.98  & 5.40 & 135.30 & 18.83   &  2.18  & 0.44 & 0.0825 &  11.7 $\pm$  0.12  & 17.1 & 9.40$\pm$ 0.05 & SG16 \\
27 & NGC 4473  & $\rm 3.6\,\mu m$ & (0.6, 3.26)  & E  & 2.30 & 45.90 & 17.93   & 2.54  & 2.90 & 36.90 & 18.10   &  2.47  & 0.46 & 0.0739 &  10.64 $\pm$  0.26  & 15.3 & 8.08$\pm$ 0.36 & SG16 \\
28 & NGC 4486\tablenotemark{a}   & $\rm 3.6\,\mu m$ & (0.6, 3.26)  & E  & 10.00 & 203.00 & 19.87   & 1.77  & 5.90 & 87.10 & 18.26   & 2.41  & 0.33 & 0.0753 &  11.49 $\pm$  0.26  &16.8 & 9.81$\pm$ 0.05 & SG16 \\
29 & NGC 4564  & $\rm 3.6\,\mu m$ & (0.6, 3.26)  & S0  & 2.60 & 5.00 & 15.23   & 3.62  & 3.00 & 6.00 & 15.65   &  3.45 & 0.45 & 0.0705 &  10.01 $\pm$  0.12  & 14.6 & 7.78$\pm$ 0.06 & SG16 \\
30 & NGC 4596  & $\rm 3.6\,\mu m$ & (0.6, 3.26)  & SB0  & 2.70 & 6.60 & 15.93   &  3.34 & 3.00 & 9.00 & 16.44   & 3.14  & 0.44 & 0.082 &  10.18 $\pm$  0.12  & 17.0 & 7.90$\pm$ 0.20 & SG16 \\
31 & NGC 4621  & $\rm 3.6\,\mu m$ & (0.6, 3.26)  & E  & 5.50 & 48.00 & 18.02   & 2.51  & 8.80 & 90.90 & 19.67   &  1.85  & 0.87 & 0.0859 &  11.16 $\pm$  0.12  & 17.8 & 8.59$\pm$ 0.05 & SG16 \\
32 & NGC 4697  & $\rm 3.6\,\mu m$ & (0.6, 3.26)  & E  & 7.20 & 239.30 & 20.62   & 1.47  & 6.70 & 226.40 & 20.90   &  1.35 & 0.53 & 0.0551 &  11.01 $\pm$  0.33  & 11.4 &8.26$\pm$ 0.05 & SG16  \\
33 & NGC 4889\tablenotemark{a}   & $\rm 3.6\,\mu m$ & (0.6, 3.26)  & E & 8.10 & 119.70 & 21.01   & 1.33  & 6.80 & 60.80 & 20.11   & 1.69  & 0.48 & 0.4863 &  12.14 $\pm$  0.12  &  103.2 & 10.32$\pm$ 0.44 & SG16 \\
34 & NGC 5077\tablenotemark{a}   & $\rm 3.6\,\mu m$ & (0.6, 3.26)  & E & 4.20 & 23.50 & 17.67   & 2.65  & 5.70 & 23.00 & 18.01   & 2.52  & 0.68 & 0.1975 &  11.28 $\pm$  0.12  & 41.2 & 8.87$\pm$ 0.22 & SG16 \\
35 & NGC 5128   & $\rm 3.6\,\mu m$ & (0.6, 3.26)  & $\rm S0^*$  & 1.20 & 61.30 & 15.73   & 3.42  & 2.20 & 60.80 & 16.01   & 3.30  & 0.42 & 0.0185 &  10.64 $\pm$  0.33  & 3.8  & 7.65$\pm$ 0.13 & SG16 \\
36 & NGC 5576  & $\rm 3.6\,\mu m$ & (0.6, 3.26)  & E    & 3.30 & 61.50 & 19.41   & 1.95 & 3.70 & 49.30 & 19.34   & 1.98  & 0.49 & 0.1194 &  10.87$\pm$  0.12  & 24.8 & 8.20$\pm$ 0.10 & SG16 \\
37 & NGC 5846\tablenotemark{a}  & $\rm 3.6\,\mu m$  & (0.6, 3.26) & E  & 6.40 & 105.10 & 19.67   & 1.85  & 5.70 & 83.40 & 19.28   & 2.01  & 0.48 & 0.1165 &  11.42 $\pm$  0.26  & 24.2  & 9.04$\pm$ 0.05 & SG16 \\
38 & NGC 6251\tablenotemark{a}  & $\rm 3.6\,\mu m$  & (0.6, 3.26) & E & 6.80 & 41.70 & 19.82   & 1.81  & 5.60 & 30.10 & 19.31   & 2.01  & 0.44 & 0.4927 &  11.82 $\pm$  0.12  & 104.6 & 8.77$\pm$ 0.16 & SG16 \\
39 & NGC 7619\tablenotemark{a}  & $\rm 3.6\,\mu m$  & (0.6, 3.26) & E & 5.30 & 63.20 & 19.53   &  1.91 & 5.20 & 58.00 & 19.55   & 1.90  & 0.51 & 0.2461 &  11.64 $\pm$ 0.26  & 51.5 & 9.40$\pm$ 0.09 & SG16 \\
40 & NGC 7768\tablenotemark{a}   & $\rm 3.6\,\mu m$  & (0.6, 3.26) & E  & 8.40 & 92.90 & 21.37   & 1.19  & 6.70 & 42.10 & 20.15   & 1.68  & 0.45 & 0.5301 &  11.89 $\pm$  0.26  & 112.8 & 9.11$\pm$ 0.15 & SG16 \\
41 & NGC 1271  & H (HST)  & (1.4, 3.33) & ES  & 4.26 & 3.25 & 16.75   & 3.43  & 4.16 & 3.07 & 16.79   & 3.41  & 0.46 & 0.3794 &  10.95$\pm$  0.1  & 80.0 & 9.48$\pm$ 0.16 & GCS16a \\
42 & NGC 1277  & V (HST)  & (11.65, 4.82) & ES & 5.34 & 6.00 & 20.73   & 3.35 & 5.63 & 5.60 & 21.05   & 3.22  & 0.56 & 0.3445 &  11.43 $\pm$ 0.1  & 72.5 & 9.08$\pm$ 0.12 & GDS16b \\
43 & A1836 BCG\tablenotemark{a}  & $\rm K_{s} $  & (0.7, 5.08)   & E  & 4.10 & 23.99 & 19.36   & 2.80 & 3.47 & 14.75 & 20.66   &  2.28  & 0.39 & 0.7335 &  11.70 $\pm$  0.12  &  158 & 9.59$\pm$ 0.06 & SGD19a \\
44 & A3565 BCG  & $\rm 3.6\,\mu m$  & (0.6, 6.02) & E  & 3.85 & 43.21 & 21.29   & 2.31 & 3.82 & 41.10 & 21.28   &  2.31 & 0.45 & 0.1951 &  11.47$\pm$  0.26  & 40.7 & 9.04$\pm$ 0.09 & SGD19a \\
45 & NGC 0307  & $\rm r^{'} (SDSS)$  & (2.8, 4.65) & SAB0 & 3.33 & 3.00 & 18.82   & 3.42  & 3.76 & 3.33 & 19.44   &  3.17 & 0.49 & 0.2523 &  10.43 $\pm$  0.33  & 52.8  & 8.34$\pm$ 0.13 & SGD19a \\
46 & NGC 0404   & $\rm 3.6\,\mu m$  & (0.6, 6.02) & S0 & 0.93 & 3.99 & 18.58   & 3.39 & 0.90 & 3.89 & 18.59   & 3.38 & 0.20 & 0.0148 &  7.96$\pm$ 0.27  & 3.1 & 4.85$\pm$ 0.13 & SGD19a \\
47 & NGC 0524\tablenotemark{a}   & $\rm 3.6\,\mu m$  & (0.6, 6.02) & SA0(rs)  & 2.29 & 8.79 & 18.67   & 3.35 & 2.16 & 8.35 & 18.57   & 3.39  & 0.34 & 0.1122 &  10.57$\pm$  0.26  & 23.3 & 8.92$\pm$ 0.10 & SGD19a \\
48 & NGC 1194\tablenotemark{a}   & $\rm 3.6\,\mu m$  & (0.6, 6.02) & $\rm S0^*$ & 3.76 & 3.52 & 18.34   & 3.49  & 3.91 & 3.56 & 18.56   & 3.40  & 0.47 & 0.2542 &  10.71$\pm$  0.33  & 53.2  & 7.81$\pm$ 0.04 & SGD19a \\
49 & NGC 1275  & $\rm 3.6\,\mu m$   & (0.6, 6.02) & E & 4.78 & 70.69 & 22.62   & 1.78 & 4.31 & 53.6 & 22.27   & 1.92 & 0.44 & 0.3464 &  11.84 $\pm$  0.26  & 72.9 & 8.90$\pm$ 0.20 & SGD19a \\
50 & NGC 1374   & $\rm 3.6\,\mu m$   & (0.6, 6.02) & S0 & 1.68 & 12.56 & 19.65   & 2.96 & 1.65 & 11.74 & 19.62   &  2.97 & 0.29 & 0.0926 &  10.22 $\pm$ 0.26  & 19.2 & 8.76$\pm$ 0.05 & SGD19a \\
51 & NGC 1407\tablenotemark{a}  & $\rm 3.6\,\mu m$  & (0.6, 6.02) & E & 3.95 & 49.67 & 20.89   &  2.46 & 3.89 & 47.29 & 20.87   &  2.47 & 0.45 & 0.1349 &  11.46 $\pm$  0.27  & 28 & 9.65$\pm$ 0.08 & SGD19a \\
52 & NGC 1550\tablenotemark{a}   & $\rm K_{s}$  & (0.7, 5.08) & E & 7.50 & 24.80 & 20.87   & 2.17 & 7.48 & 24.8 & 21.16   &  2.05 & 0.58 & 0.2465 &  11.13 $\pm$  0.12  & 51.6 & 9.57$\pm$ 0.06 & SGD19a \\
53 & NGC 1600\tablenotemark{a}   & $\rm 3.6\,\mu m$  & (0.6, 6.02) & E & 7.14 & 76.57 & 22.66   & 1.77  & 5.08 & 49.58 & 21.98   & 2.04  & 0.38 & 0.3048 &  11.82 $\pm$  0.12  & 64 & 10.23$\pm$ 0.05 & SGD19a \\
54 & NGC 2787  & $\rm 3.6\,\mu m$  & (0.6, 6.02) & SB0(r)  & 1.36 & 4.06 & 17.32   &  3.89 & 1.27 & 2.88 & 17.08   & 3.99  & 0.25 & 0.0353 &  9.13 $\pm$ 0.26  & 7.3 & 7.60$\pm$ 0.06 & SGD19a \\
55 & NGC 3665  & $\rm 3.6\,\mu m$  & (0.6, 6.02) & S0 & 2.76 & 14.44 & 19.28   & 3.11 & 2.74 & 12.78 & 19.34   &  3.09 & 0.39 & 0.1666 &  11.03 $\pm$ 0.26  & 34.7 & 8.76$\pm$ 0.10 & SGD19a \\
56 & NGC 3923\tablenotemark{a}  & $\rm 3.6\,\mu m$  & (0.6, 6.02) & E  & 4.78 & 85.97 & 21.50   & 2.22 & 4.77 & 78.78 & 21.55   & 2.20  & 0.5 & 0.1006 &  11.4 $\pm$  0.15  & 20.9  & 9.45$\pm$ 0.13 & SGD19a \\
57 & NGC 4026   & $\rm 3.6\,\mu m$  & (0.6, 6.02) & SB0  & 3.45 & 3.10 & 16.14   & 4.36 & 3.98 & 2.35 & 16.04   & 4.40  & 0.51 & 0.0638 &  10.11 $\pm$  0.33  & 13.2 & 8.26$\pm$ 0.11 & SGD19a \\
58 & NGC 4339   & $\rm 3.6\,\mu m$  & (0.6, 6.02) & S0  & 1.46 & 6.64 & 19.22   & 3.13 & 1.40 & 6.42 & 19.21   & 3.14  & 0.27 & 0.0772 &  9.67 $\pm$  0.26  & 16.0  & 7.63$\pm$ 0.33 & SGD19a \\
59 & NGC 4342   & $\rm 3.6\,\mu m$  & (0.6, 6.02) & ES  & 3.48 & 4.22 & 18.48   & 3.43 & 3.99 & 4.69 & 19.23   & 3.13  & 0.51 & 0.1108 &  9.94 $\pm$ 0.25  & 23.0 & 8.65$\pm$ 0.18 & SGD19a \\
60 & NGC 4350   & $\rm 3.6\,\mu m$  & (0.6, 6.02) & EBS  & 4.30 & 18.84 & 20.8   & 2.50 & 3.97 & 19.45 & 20.75   &  2.52 & 0.44 & 0.0811 &  10.28 $\pm$ 0.26  & 16.8 & 8.86$\pm$ 0.41 & SGD19a \\
61 & NGC 4371   & $\rm 3.6\,\mu m$  & (0.6, 6.02) & SB(r)0 & 2.83 & 8.46 & 19.49   & 3.02 & 3.19 & 8.90 & 19.91   & 2.85  & 0.45 & 0.0816 &  9.89 $\pm$ 0.26  & 16.9 & 6.85$\pm$ 0.08 & SGD19a \\
62 & NGC 4429    & $\rm 3.6\,\mu m$  & (0.6, 6.02) & SB(r)0 & 2.56 & 16.42 & 19.05   & 3.20 & 2.31 & 11.29 & 18.78   & 3.31  & 0.34 & 0.0796 &  10.46$\pm$  0.26  & 16.5 & 8.18$\pm$ 0.09 & SGD19a \\
63 & NGC 4434   & $\rm 3.6\,\mu m$  & (0.6, 6.02) & S0 & 2.68 & 4.94 & 19.06   & 3.20 & 2.93 & 5.31 & 19.31   & 3.10  & 0.43 & 0.1079 &  9.91 $\pm$ 0.26  & 22.4 & 7.85$\pm$ 0.17 & SGD19a \\
64 & NGC 4486B  & $\rm r^{'} (SDSS)$  & (2.8, 4.65) & E   & 2.63 & 2.60 & 18.20   & 3.66  & 2.74 & 2.53 & 18.30   & 3.62  & 0.4 & 0.0739 &  9.47 $\pm$ 0.33  & 15.3  & 8.76$\pm$ 0.24 & SGD19a \\
65 & NGC 4526   & $\rm 3.6\,\mu m$   & (0.6, 6.02) & S0  & 2.28 & 13.01 & 18.19   & 3.54 & 2.96 & 14.88 & 18.98   & 3.23  & 0.47 & 0.0816 &  10.7 $\pm$  0.26  & 16.9 & 8.67$\pm$ 0.05 & SGD19a \\
66 & NGC 4552   & $\rm 3.6\,\mu m$   & (0.6, 6.02) & E & 5.42 & 83.62 & 22.12   & 1.97 & 5.36 & 71.5 & 21.92   &  2.05 & 0.52 & 0.0719 &  10.88 $\pm$  0.25  & 14.9 & 8.67$\pm$ 0.05 & SGD19a \\
67 & NGC 4578   & $\rm 3.6\,\mu m$   & (0.6, 6.02)  & S0(r) & 2.30 & 7.82 & 19.26   &  3.11 & 1.99 & 6.32 & 19.14   & 3.16  & 0.31 & 0.0787 &  9.77 $\pm$  0.26  & 16.3  & 7.28$\pm$ 0.35 & SGD19a \\
58 & NGC 4649\tablenotemark{a}  & $\rm 3.6\,\mu m$   & (0.6, 6.02) & E  & 4.96 & 93.03 & 21.01   & 2.42  & 5.21 & 80.59 & 20.98   &  2.43 & 0.54 & 0.0791 &  11.44 $\pm$  0.12  & 16.4  & 9.67$\pm$ 0.10 & SGD19a \\
69 & NGC 4742  & $\rm 3.6\,\mu m$   & (0.6, 6.02) & S0  & 2.62 & 3.37 & 17.13   & 3.97 & 3.20 & 3.41 & 17.62   & 3.77  & 0.48 & 0.0748 &  9.87 $\pm$ 0.26  & 15.5 & 7.15$\pm$ 0.18 & SGD19a \\
70 & NGC 4751\tablenotemark{a}  & $\rm K_{s}$  & (0.7, 5.08) & S0  & 3.79 & 12.47 & 18.74   & 3.02  & 3.25 & 5.48 & 17.76   & 3.41  & 0.38 & 0.1295 &  10.49 $\pm$  0.26  & 26.9  & 9.15$\pm$ 0.05 & SGD19a \\
71 & NGC 4762  & $\rm 3.6\,\mu m$   & (0.6, 6.02) &  SB0 & 2.36 & 4.39 & 17.89   & 3.66 & 1.85 & 2.24 & 17.09   & 3.98  & 0.29 & 0.1089 &  9.97 $\pm$  0.28  & 22.6 & 7.36$\pm$ 0.15 & SGD19a \\
72 & NGC 5018  & $\rm 3.6\,\mu m$    & (0.6, 6.02) & $\rm S0^*$ & 2.64 & 8.29 & 18.4   & 3.46 & 2.51 & 6.20 & 18.22   & 3.54  & 0.36 & 0.1944 &  10.98 $\pm$  0.27  & 40.6 & 8.02$\pm$ 0.09 & SGD19a \\
73 & NGC 5252  & $\rm 3.6\,\mu m$   & (0.6, 6.02) & S0    & 3.08 & 2.07 & 17.82   & 3.71 & 2.95 & 1.47 & 17.46   &  3.85  & 0.39 & 0.4569 &  10.85 $\pm$  0.26  & 96.8 & 9.00$\pm$ 0.40 & SGD19a \\
74 & NGC 5328\tablenotemark{a}  & $\rm K_{s} $  & (0.7, 5.08) & E  & 6.58 & 25.11 & 20.29   & 2.40 & 5.21 & 22.46 & 20.26   &  2.42 & 0.42 & 0.3053 &  11.49 $\pm$  0.12  & 64.1 & 9.67$\pm$ 0.15 & SGD19a \\
75 & NGC 5419\tablenotemark{a}   & $\rm 3.6\,\mu m$   & (0.6, 6.02) & E  & 2.42 & 17.24 & 19.80   &  2.91 & 2.62 & 16.83 & 20.01   &  2.82 & 0.40 & 0.2683 &  11.45 $\pm$  0.12  & 56.2 & 9.86$\pm$ 0.14 & SGD19a \\
76 & NGC 5516\tablenotemark{a}   & $\rm K_{s}$  & (0.7, 5.08) & E & 5.99 & 51.27 & 21.54  & 1.90  & 5.32 & 32.30 & 20.86   &  2.18 & 0.47 & 0.2788 &  11.44 $\pm$  0.12  & 58.4 & 9.52$\pm$ 0.06 & SGD19a \\
77 & NGC 5813\tablenotemark{a}   & $\rm 3.6\,\mu m$   & (0.6, 6.02) & S0 & 4.02 & 18.54 & 20.74   & 2.53 & 3.65 & 14.16 & 19.81   & 2.90  & 0.42 & 0.1504 &  10.86 $\pm$  0.26  & 31.3 & 8.83$\pm$ 0.06 & SGD19a \\
78 & NGC 5845  & $\rm 3.6\,\mu m$   & (0.6, 6.02) & ES  & 3.33 & 6.02 & 19.16   & 3.16 & 3.27 & 5.29 & 19.10   & 3.18  & 0.42 & 0.1213 &  10.12 $\pm$  0.26  & 25.2  & 8.41$\pm$ 0.22 & SGD19a \\
79 & NGC 6086\tablenotemark{a}  & $\rm r^{'} (SDSS)$   & (2.8, 4.65) & E & 4.37 & 13.89 & 21.68   & 2.30 & 4.20 & 12.41 & 21.78   & 2.26  & 0.46 & 0.6441 &  11.52 $\pm$  0.26  & 138  & 9.57$\pm$ 0.17 & SGD19a \\
80 & NGC 6861  & $\rm 3.6\,\mu m$   & (0.6, 6.02) & ES  & 3.07 & 18.69 & 19.35   & 3.08 & 3.52 & 20.13 & 20.17   &  2.75 & 0.48 & 0.1314 &  10.94$\pm$ 0.29  & 27.3  & 9.30$\pm$ 0.08 & SGD19a \\
81 & NGC 7052\tablenotemark{a}   & $\rm 3.6\,\mu m$   & (0.6, 6.02) & E  & 3.20 & 34.42 & 21.84   & 2.36  & 3.46 & 20.04 & 20.82   & 2.50  & 0.46 & 0.3161 &  11.46 $\pm$  0.12  & 66.4 & 8.57$\pm$ 0.23 & SGD19a \\
82 & NGC 7332  & $\rm 3.6\,\mu m$   & (0.6, 6.02) & EB0   & 1.78 & 2.87 & 16.78   & 4.11 & 2.15 & 2.43 & 16.92   & 4.05  & 0.37 & 0.1198 &  10.22 $\pm$  0.34  & 24.9 & 7.11$\pm$ 0.20 & SGD19a \\
83 & NGC 7457   & $\rm 3.6\,\mu m$   & (0.6, 6.02) & S0   & 2.63 & 6.31 & 19.62   & 2.97 & 2.84 & 6.51 & 19.98   & 2.83  & 0.42 & 0.0676 &  9.40 $\pm$  0.26  & 14  & 7.00$\pm$ 0.30 & SGD19a \\
84 & Circinus   & $\rm 3.6\,\mu m$   & (0.45, 6.02) & SABb   & 2.21 & 33.26 & 18.29   & 3.37 & 1.80 & 23.13 & 18.06   & 3.47  & 0.29 & 0.0204 &  10.12 $\pm$  0.2  & 4.2  & 6.25$\pm$ 0.11 & DGC19a \\
85 & ESO558-G009   & I (HST)   & (1.88, 4.52) & Sbc   & 1.28 & 0.62 & 18.17   &  3.47 & 1.63 & 0.68 & 18.62   & 3.29  & 0.31 & 0.542 &  9.89 $\pm$  0.11  & 115.4 & 7.26$\pm$ 0.04  & DGC19a \\
86 & IC 2560  & $\rm 3.6\,\mu m$   & (0.45, 6.02) & SBb   & 2.27 & 7.15 & 19.64   &  2.84 & 0.68 & 3.92 & 19.07   & 3.07  & 0.15 & 0.149 &  9.63 $\pm$  0.39  & 31.0 & 6.49$\pm$ 0.20 & DGC19a  \\
87 & J0437+2456  &  I (HST)   & (1.88, 4.52) & SB    & 1.73 & 1.22 & 19.42   &  2.96 & 1.97 & 0.87 & 19.40   &  2.97 & 0.35 & 0.343 &  9.90 $\pm$  0.2  & 72.8 & 6.51$\pm$ 0.05 & DGC19a  \\
88  & Milky Way  & $\rm 2.4\,\mu m$   & ...  & SBbc    & 1.32 &  $\rm 7.49 \degree$  & ...     &  ... & 1.32 & $\rm 5.85 \degree$ & ...    & ... & ... & 0.38E-4 &  9.96 $\pm$  0.05  & 7.86E-3 & 6.6$\pm$ 0.02 & GD07 \\
89 & Mrk 1029   &  I (HST)   & (1.88, 4.52) & S   & 1.15 & 0.47 & 16.53   & 4.13 & 1.07 & 0.28 & 16.29   &  4.23 & 0.23 & 0.6392 &  9.90 $\pm$ 0.11  & 136.9  & 6.33$\pm$ 0.12 & DGC19a \\
90 & NGC 0224   & $\rm 3.6\,\mu m$   & (0.45, 6.02) & SBb   & 2.20 & 418.6 & 19.58   & 2.87 & 1.30 & 173.6 & 18.41   &  3.33 & 0.22 & 0.0036 &  10.11 $\pm$  0.09  & 0.8  & 8.15$\pm$ 0.16 & DGC19a \\
91 & NGC 0253   & $\rm 3.6\,\mu m$   & (0.45, 6.02) & SABc   & 2.53 & 55.55 & 19.22   &  3.00 & 2.33 & 27.89 & 18.82   &  3.16 & 0.34 & 0.0168 &  9.76 $\pm$  0.09  & 3.5 & 7.00$\pm$ 0.30 & DGC19a \\
92 & NGC 1068   & $\rm K_{s}$   & (0.62, 5.08) & SBb   & 0.71 & 10.52 & 16.17   &  3.99 & 0.87 & 8.29 & 16.14   & 4.00  & 0.20 & 0.0488 &  10.27 $\pm$  0.24  & 10.1  & 6.75$\pm$ 0.08 & DGC19a \\
93 & NGC 1097   & $\rm 3.6\,\mu m$   & (0.45, 6.02) & SBb   & 1.95 & 15.72 & 18.71   &  3.21 & 1.52 & 11.39 & 18.27   &  3.39 & 0.26 & 0.1199 &  10.83 $\pm$  0.2  & 24.9 & 8.38$\pm$ 0.04 & DGC19a \\
94 & NGC 1300    & $\rm 3.6\,\mu m$   & (0.45, 6.02) & SBbc  & 4.20 & 24.37 & 21.97   &  1.91 & 2.83 & 7.39 & 19.99   &  2.70 & 0.31 & 0.07 &  9.42 $\pm$  0.25  & 14.5 & 7.71$\pm$ 0.16 & DGC19a \\
95 & NGC 1320    & $\rm 3.6\,\mu m$   & (0.45, 6.02) & Sa  & 3.08 & 3.35 & 17.93   &  3.53 & 2.87 & 2.23 & 17.40   & 3.74  & 0.38 & 0.1809 &  10.25 $\pm$  0.4  & 37.7 & 6.78$\pm$ 0.29 & DGC19a \\
96 & NGC 1398    & $\rm 3.6\,\mu m$   & (0.45, 6.02) & SBab  & 3.44 & 17.53 & 19.75   &  2.80 & 3.00 & 10.38 & 19.04   & 3.08  & 0.38 & 0.1194 &  10.57 $\pm$  0.2  & 24.8 & 8.03$\pm$ 0.11 & DGC19a \\
97 & NGC 2273    &  I (HST)   & (1.88, 4.52) & SBa  & 2.24 & 2.99 & 18.13   &  3.47 & 2.49 & 3.15 & 18.52   & 3.31 & 0.39 & 0.1519 &  9.98 $\pm$  0.2  & 31.6 & 6.97$\pm$ 0.09 & DGC19a \\
98 & NGC 2960    & $\rm 3.6\,\mu m$   & (0.45, 6.02) & $\rm Sa^*$  & 2.59 & 2.35 & 18.04   & 3.49 & 2.86 & 2.19 & 18.30   & 3.39  & 0.42 & 0.338 &  10.44 $\pm$  0.36  & 71.1 & 7.06$\pm$ 0.17 & DGC19a \\
99 & NGC 2974    & $\rm 3.6\,\mu m$   & (0.45, 6.02) & SB  & 1.56 & 9.21 & 18.49   & 3.30 & 1.17 & 6.53 & 18.12   & 3.45  & 0.23 & 0.1036 &  10.23 $\pm$  0.13  & 21.5 & 8.23$\pm$ 0.07 & DGC19a \\
100 & NGC 3031\tablenotemark{b}   & $\rm 3.6\,\mu m$   & (0.45, 6.02) & Sab  & 2.81 & 36.19 & 18.34   & 3.35 & 3.46 & 42.98 & 18.93   & 3.12  & 0.50 & 0.0169 &  10.16 $\pm$  0.11  & 3.5  & 7.83$\pm$ 0.09 & DGC19a \\
101 & NGC 3079   & $\rm 3.6\,\mu m$   & (0.45, 6.02) & SBcd  & 0.52 & 5.91 & 16.79   & 3.98  & 0.58 & 4.35 & 17.13   & 3.84  & 0.16 & 0.0796 &  9.92 $\pm$  0.25  & 16.5 & 6.38$\pm$ 0.12 & DGC19a \\
102 & NGC 3227   & $\rm 3.6\,\mu m$   & (0.45, 6.02) & SABa   & 2.60 & 17.91 & 20.26   & 2.59 & 1.90 & 8.34 & 19.32   & 2.97  & 0.28 & 0.1017 &  10.04 $\pm$  0.17  & 21.1 & 7.88$\pm$ 0.14 & DGC19a \\
103 & NGC 3368    & $\rm 3.6\,\mu m$   & (0.45, 6.02) & SABa  & 1.19 & 5.98 & 17.07   & 3.86  & 1.00 & 4.83 & 16.92   &  3.92 & 0.21 & 0.0517 &  9.81 $\pm$  0.1  & 10.7 & 6.89$\pm$ 0.09 & DGC19a \\
104 & NGC 3393    &  I (HST)   & (1.88, 4.52) & SBa  & 1.14 & 1.64 & 17.27   & 3.82  & 1.36 & 1.77 & 17.63   &  3.68 & 0.27 & 0.2664 &  10.23 $\pm$  0.12  & 55.8 & 7.49$\pm$ 0.05 & DGC19a \\
105 & NGC 3627    & $\rm 3.6\,\mu m$    & (0.45, 6.02) & SBb & 3.17 & 11.07 & 18.44   & 3.32  & 2.10 & 3.92 & 16.98   & 3.90  & 0.28 & 0.0512 &  9.74 $\pm$  0.2  & 10.6 & 6.95$\pm$ 0.05 & DGC19a \\
106 & NGC 4151   & $\rm 3.6\,\mu m$   & (0.45, 6.02) & SABa   & 2.24 & 6.23 & 17.75   & 3.59  & 1.85 & 6.00 & 17.77   &  3.59 & 0.29 & 0.0916 &  10.27 $\pm$  0.15  & 19.0 &  7.68$\pm$ 0.37 & DGC19a \\
107 & NGC 4258   & $\rm 3.6\,\mu m$   & (0.45, 6.02) & SABb   & 3.21 & 41.8 & 20.14   &  2.64 & 2.60 & 26.4 & 19.73   & 2.80  & 0.34 & 0.0368 &  10.05 $\pm$  0.18  & 7.6 & 7.60$\pm$ 0.01 & DGC19a \\
108 & NGC 4303   & $\rm 3.6\,\mu m$    & (0.45, 6.02) & SBbc  & 1.02 & 2.28 & 16.51   &  4.09 & 0.90 & 2.16 & 15.78   &  4.38 & 0.20 & 0.0594 &  9.42 $\pm$  0.1  & 12.3 & 6.58$\pm$ 0.17 & DGC19a \\
109 & NGC 4388    & $\rm 3.6\,\mu m$   & (0.45, 6.02) & SBcd  & 0.89 & 21.68 & 21.68   &  2.02 & 1.15 & 14.3 & 19.82   & 2.77  & 0.24 & 0.0859 &  10.07 $\pm$  0.22  & 17.8 & 6.90$\pm$ 0.11 & DGC19a \\
110 & NGC 4501    & $\rm 3.6\,\mu m$   & (0.45, 6.02) & Sb  & 2.33 & 21.22 & 19.53   &  2.88 & 2.83 & 20.35 & 19.91   &  2.73 & 0.44 & 0.0541 &  10.11 $\pm$  0.16  & 11.2  & 7.13$\pm$ 0.08 & DGC19a \\
111 & NGC 4594    & $\rm 3.6\,\mu m$   & (0.45, 6.02) & Sa  & 6.14 & 44.94 & 19.38   &  2.94 & 4.24 & 41.36 & 19.46   &  2.91 & 0.35 & 0.0462 &  10.81 $\pm$ 0.2  & 9.6 & 8.81$\pm$ 0.03 & DGC19a \\
112 & NGC 4699   & $\rm 3.6\,\mu m$   & (0.45, 6.02) & SABb   & 5.35 & 24.44 & 19.51   &  2.89 & 6.77 & 29.75 & 20.31   & 2.57  & 0.69 & 0.1141 &  11.12 $\pm$ 0.26  & 23.7 & 8.34$\pm$ 0.10 & DGC19a \\
113 & NGC 4736\tablenotemark{b}   & $\rm 3.6\,\mu m$   & (0.45, 6.02) & Sab   & 0.93 & 9.79 & 16.17   &  4.22 & 1.03 & 9.65 & 16.31   & 4.17  & 0.22 & 0.0214 &  9.89 $\pm$ 0.09  & 4.4 & 6.78$\pm$ 0.10 & DGC19a \\
114 & NGC 4826   & $\rm 3.6\,\mu m$   & (0.45, 6.02) & Sab  & 0.73 & 13.89 & 17.86   &  3.55 & 0.76 & 11.93 & 17.98   &  3.50 & 0.18 & 0.0269 &  9.55 $\pm$  0.22  & 5.6 & 6.07$\pm$ 0.15 & DGC19a \\
115 & NGC 4945\tablenotemark{b}  & $\rm K_{s}$   & (0.62, 5.08) & Sc   & 3.40 & 26.33 & 18.48   &  3.06 & 3.19 & 13.93 & 17.99   & 3.26  & 0.40 & 0.018 &  9.39 $\pm$  0.19  & 3.7 & 6.15$\pm$ 0.30 & DGC19a \\
116 & NGC 5055   & $\rm 3.6\,\mu m$   & (0.45, 6.02) & Sbc   & 2.02 & 55.12 & 20.09   & 2.66  & 1.76 & 43.52 & 19.85   & 2.75  & 0.29 & 0.0429 &  10.49 $\pm$  0.11  & 8.9 & 8.94$\pm$ 0.10 & DGC19a \\
117 & NGC 5495    &  I (HST)   & (1.88, 4.52) & SBc  & 2.60 & 3.75 & 20.21   &  2.65 & 2.46 & 3.99 & 20.23   & 2.64  & 0.36 & 0.4767 &  10.54 $\pm$  0.12  & 101.1 & 7.04$\pm$ 0.08 & DGC19a \\
118 & NGC 5765b   &  I (HST)   &(1.88, 4.52) & SABb    & 1.46 & 1.11 & 18.72   &  3.25 & 1.51 & 1.00 & 18.83   &  3.21 & 0.28 & 0.6257 &  10.04 $\pm$  0.13  & 133.9  & 7.72$\pm$ 0.05 & DGC19a \\
119 & NGC 6264   &  I (HST)   & (1.88, 4.52) & SBb   & 1.04 & 1.13 & 19.23   & 3.06  & 1.35 & 1.05 & 19.37   &  3.00 & 0.27 & 0.7152 &  10.01 $\pm$  0.15  & 153.9 & 7.51$\pm$ 0.06 & DGC19a \\
120 & NGC 6323   &  I (HST)   & (1.88, 4.52) & SBab   & 1.60 & 1.53 & 20.39   &  2.58 & 1.15 & 1.71 & 19.98   & 2.75  & 0.22 & 0.5488 &  9.86 $\pm$  0.31  & 116.9 & 7.02$\pm$ 0.14 & DGC19a \\
121 & NGC 7582   & $\rm 3.6\,\mu m$   & (0.45, 6.02) & SBab   & 2.20 & 5.33 & 17.04   &  3.88 & 2.21 & 4.55 & 17.66   &  3.63 & 0.35 & 0.0959 &  10.15 $\pm$  0.2  & 19.9 & 7.67$\pm$ 0.09 & DGC19a \\
122 & UGC 3789    &  I (HST)   & (1.88, 4.52) & SABa  & 2.37 & 1.60 & 18.38   &  3.37 & 2.67 & 3.11 & 19.03   & 3.11  & 0.41 & 0.2372 &  10.18 $\pm$  0.14  & 49.6 & 7.06$\pm$ 0.05 & DGC19a \\
123 & UGC 6093    &  I (HST)   & (1.88, 4.52) & SBbc  & 1.55 & 1.84 & 19.27   & 3.04 & 1.41 & 1.27 & 18.87   & 3.20  & 0.26 & 0.7103 &  10.35 $\pm$  0.14  & 152.8 & 7.41$\pm$ 0.03 & DGC19a 
\label{Total Sample}
\enddata
\tablecomments{Column: (1) Galaxy name. 
(2) Wavelength-band ($\lambda$) of the image used in parent studies (Column 18). Images for the first 41 galaxies were calibrated to Vega magnitude system and the images of the remaining galaxies (except the Milky Way)  were calibrated to AB magnitude system. 
(3) The stellar mass-to-light ratio  ($\Upsilon_{\lambda}$) and the absolute magnitude of the Sun ($\mathfrak{M}_{\odot,\lambda}$) used to obtain the bulge stellar mass. \citet{Davis:2018:a} used a reduced (by $\rm 25\%$) stellar mass-to-light ratio for their LTGs observed at $3.6\,\mu \rm m$-band following \citet{Querejeta:2015} who reported on the dust glow at $3.6\,\mu \rm m$ in LTGs.
(4) Galaxy morphology based on the multi-component decomposition of the galaxy light performed in studies listed in Column 18. Galaxy mergers are highlighted with an $*$.
(5) Bulge major-axis S\'ersic index parameter.
(6) Bulge major-axis effective half-light radius.
(7) Bulge surface brightness at the corresponding major-axis half-light radius listed in column 6.
(8) Logarithm of the  bulge intensity at the major-axis half-light radius in the units of $\rm M_\odot / pc^2$, calculated using $\rm [(\mu_e - Dist. Mod. - \mathfrak{M}_{\odot, \lambda}-2.5 \log(1/scale^2)-2.5log(\Upsilon_\lambda))/(-2.5)]$ \citep[][their Equation 10]{Graham:Merritt:2006}.
(9)-(12) Similar to columns (5)-(8), but obtained from an independent multi-component decomposition of the galaxy light profile along the equivalent-axis ($\rm R_{eq} = \sqrt{R_{maj}* R_{min}} $).
(13) Concentration index (C) calculated using the equivalent-axis bulge S\'ersic index and  Equation 6 from \citet{Trujillo:Graham:Caon:2001} using $\alpha =1/3$.
(14) Physical scale in $\rm kpc \, arcsec^{-1}$, assuming cosmological parameters from \citet{Planck:Collaboration:2018}.
(15) Logarithm of the spheroid stellar mass in units of solar mass.
(16) Galaxy  distance in megaparsec.
(17) Logarithm of the directly-measured black hole mass in units of solar mass. 
(18) Parent studies which performed multi-component decompositions to obtain the bulge parameters. Where SG16=\citet{Savorgnan:Graham:2016:I}, SGD19a=\citet{Sahu:2019:I}, DGC19a=\citet{Davis:2018:a}, GCS16a=\citet{Graham:Ciambur:Savorgnan:2016} , GDS16b=\citet{Graham:Durr:Savorgnan:2016}, and GD07=\citet{Graham:Driver:2007}. Original sources for black hole mass and distances can be found in \citet{Savorgnan:2016:Slopes}, and \citet{Sahu:2019:I} for ETGs and  \citet{Davis:2018:a} for LTGs.
\tablenotetext{a}{Galaxies with a deficit of light at their center, for whom the spheroid profile is parameterized using a core-S\'ersic function \citep{Graham:2003:CS}.}
\tablenotetext{b}{The morphology of these spiral galaxies listed on \textsc{HyperLeda} \citep{Paturel:2003} or NASA/IPAC Extragalactic Database (NED) suggests a weak bar, however, \citet{Davis:2018:a} did not find evidence of an  extended/intermediate  bar in these galaxies but rather a nuclear bar.}
}
\end{deluxetable*} 
\end{longrotatetable}
\endgroup
 \(\)

\startlongtable
\begin{deluxetable*}{lcrrccrrrr}
\tabletypesize{\footnotesize}
\tablecolumns{6}
\tablecaption{Correlations of $M_{\rm *,sph}$ and $M_{\rm BH}$ with the bulge equivalent-axis properties ($\rm n_{eq,sph},  C(1/3), and \, R_{e,sph,eq}$) calculated using  a symmetric application of the \textsc{MPFITEXY} regression\label{fit parameters3} (see Section \ref{Data})}
\tablehead{
\colhead{ \textbf{Category} } & \colhead{ \textbf{Number} } & \colhead{ \bm{$\alpha$} } & \colhead{ \bm{$\beta$ } } & \colhead{ \bm{$\epsilon$} } & \colhead{ \bm{ $\Delta_{\rm rms}$} }  \\
\colhead{} & \colhead{} & \colhead{} & \colhead{ \textbf{dex}} & \colhead{\textbf{dex}} & \colhead{\textbf{dex}}  \\
\colhead{ \textbf{(1)}} & \colhead{\textbf{(2)}} & \colhead{\textbf{(3)}} & \colhead{\textbf{(4)}} & \colhead{\textbf{(5)}} & \colhead{\textbf{(6)}} 
}
\startdata
\multicolumn{6}{c}{ $\log(M_{*,\rm sph}/{\rm M_{\sun}})=\alpha\log(\rm n_{sph,eq}/3)+\beta$} \\
\hline
ETGs & 77 & $ 3.36 \pm 0.20  $ &  $ 10.52\pm 0.04 $ &  0.30 & 0.48  \\
LTGs & 38 & $1.47\pm 0.19 $ & $10.48 \pm 0.06  $ & 0.20 & 0.29  \\
\hline
\multicolumn{6}{c}{ $\log(M_{\rm BH}/{\rm M_{\sun}})=\alpha\log(\rm n_{sph,eq}/3)+\beta$} \\
\hline
ETGs &  77 & $ 3.94\pm 0.36 $ & $8.18 \pm 0.07 $ & 0.62 & 0.73   \\
LTGs & 38  & $ 2.90\pm 0.55 $ & $ 8.00\pm 0.18 $ &  0.63 & 0.69   \\
\hline
\multicolumn{6}{c}{ $\log(M_{\rm BH}/{\rm M_{\sun}})=\alpha \,  \rm C(1/3)/0.4+\beta$} \\
\hline
ETGs &  77 & $ 8.85 \pm 0.81 $ & $8.10 \pm 0.08 $ & 0.62 & 0.72 \\
LTGs &  38 & $ 7.03 \pm 1.50 $ & $7.94 \pm 0.18 $ & 0.64 & 0.68   \\
\hline
\multicolumn{6}{c}{ $\log(M_{*,\rm sph}/{\rm M_{\sun}})=\alpha\log(\rm R_{e,sph, eq})+\beta$} \\
\hline
All Galaxies &  115 & $ 1.12 \pm 0.03 $ & $10.42 \pm 0.02 $ & 0.07 & 0.25  \\
\hline
\multicolumn{6}{c}{ $\log(M_{\rm BH}/{\rm M_{\sun}})=\alpha\log(\rm R_{e, sph,eq})+\beta$} \\
\hline
ETGs with a disk & 39 & $2.08 \pm 0.23 $ & $8.49 \pm 0.09 $ &  0.51 & 0.60  \\
ETGs without a disk & 38  & $ 2.09\pm 0.35 $ & $7.12 \pm 0.36 $ &  0.53 & 0.61  \\
ETGs & 77 & $1.30 \pm 0.09 $ & $8.10 \pm 0.06 $ &  0.56 & 0.60 \\
LTGs & 38  & $ 2.41\pm 0.29 $ & $7.79 \pm 0.10 $ &  0.51 & 0.60 \\
\enddata
\tablecomments{Columns:
(1) Subclass of galaxy.
(2) Number of galaxies in subclass.
(3) Slope of the line obtained from the \textsc{MPFITEXY(Bisector)} regression.
(4) Intercept of the line obtained from the \textsc{MPFITEXY(Bisector)} regression.
(5) Intrinsic scatter in the vertical ($\log M_{\rm *,sph}$ or $\log M_{\rm BH}$)-direction.
(6) Total root mean square (rms) scatter in the vertical direction.
}
\end{deluxetable*}
\(\)

\end{document}